\documentclass[apj,twocolumn]{openjournal}
\usepackage[utf8]{inputenc}
\usepackage{amsmath}
\usepackage{hyperref}
\usepackage{yhmath}
\newcommand{\cm}{\checkmark}
\newcommand{\crs}{\boldsymbol{\times}}
\newcommand{\mi}{\mathrm{i}}
\newcommand{\md}{\mathrm{d}}
\newcommand{\bb}[1]{\mathbf{#1}}
\usepackage{xcolor}
\usepackage[T1]{fontenc}  
\usepackage{aas_macros}

\usepackage{comment}
\usepackage{natbib}
\setcitestyle{numbers}
\begin{document}
\rightline{\scriptsize RBI-ThPhys-2025-26}

\title{The Bispectrum of Intrinsic Alignments: II. Precision Comparison Against Dark Matter Simulations}

\author{\vspace{-1.5cm}Thomas Bakx$^{\dagger,1}$}
\author{Toshiki Kurita$^{2,3}$}
\author{Alexander Eggemeier$^4$}
\author{Nora Elisa Chisari$^{1,5}$}
\author{Zvonimir Vlah$^{6,7,8}$}
\email{$\dagger$ t.j.m.bakx@uu.nl}
\affiliation{$^1$Institute for Theoretical Physics, Utrecht University, Princetonplein 5, 3584 CC, Utrecht, The Netherlands.}
\affiliation{$^2$Max-Planck-Institut für Astrophysik, Karl-Schwarzschild-Str. 1, 85748 Garching, Germany}
\affiliation{$^3$Kavli Institute for the Physics and Mathematics of the Universe (WPI), \\
The University of Tokyo Institutes for Advanced Study (UTIAS), The University of Tokyo, Chiba 277-8583, Japan}
\affiliation{$^4$Universität Bonn, Argelander-Institut für Astronomie, Auf dem Hügel 71, 53121 Bonn, Germany}
\affiliation{$^5$Leiden Observatory, Leiden University, PO Box 9513, 2300 RA Leiden, the Netherlands.}
\affiliation{$^6$Division of Theoretical Physics, Ruđer Bo\v{s}kovi\'c Institute, 10000 Zagreb, Croatia,}%
\affiliation{$^7$Kavli Institute for Cosmology, University of Cambridge, Cambridge CB3 0HA, UK}%
\affiliation{$^8$Department of Applied Mathematics and Theoretical Physics, University of Cambridge, Cambridge CB3 0WA, UK.}

\begin{abstract}
We measure three-dimensional bispectra of halo intrinsic alignments (IA) and dark matter overdensities in real space from N-body simulations for halos of mass $10^{12}-10^{12.5} M_\odot /h$. We show that their multipoles with respect to the line of sight can be accurately described by a tree-level perturbation theory model on large scales ($k\lesssim 0.11\,h$/Mpc) at $z=0$.  For these scales and in a simulation volume of 1 (Gpc/$h)^3$, we detect the bispectrum monopole $B_{\delta\delta E}^{00}$ at SNR $\sim 30$ and the two quadrupoles $B_{\delta \delta E}^{11}$ and $B_{\delta \delta E}^{20}$ at SNR $\sim 25$ and SNR $\sim 15$, respectively. We also report similar SNR for the lowest order multipoles of $B_{\delta EE}$ and $B_{EEE}$, although these are largely driven by stochastic contributions. We show that the first and second order EFT parameters are consistent with those obtained from fitting the IA power spectrum analysis at next-to-leading order, without requiring any priors to break degeneracies for the quadratic bias parameters. Moreover, the inclusion of higher multipole moments of $B_{\delta\delta E}$  greatly reduces the errors on second order bias parameters, by factors of 5 or more. The IA bispectrum thus provides an effective means of determining higher order shape bias parameters, thereby characterizing the scale dependence of the IA signal. We also detect parity-odd bispectra such as $B_{\delta \delta B}$ and $B_{\delta EB}$ at $\sim 10 \sigma$ significance or more for $k<0.15\,h$/Mpc and they are consistent with the parity-even sector. Furthermore, we check that the Gaussian covariance approximation works reasonably well on the scales we consider here. These results lay the groundwork for using the bispectrum of IA in cosmological analyses.
\end{abstract}

\maketitle
\section{Introduction}
The most recent decade has been one of many fundamental developments in cosmology, both observationally and theoretically. Notably, the advent of deep and wide galaxy surveys such as BOSS \cite{alam_boss} and DESI \cite{desi_design} has provided the scientific community orders of magnitude more data on the large-scale structure (LSS) of the Universe than what was previously available. This trend will continue well with surveys such as Euclid \cite{laureijs}, SPHEREx \cite{spherex} and LSST \cite{lsst} pushing to yet higher redshifts and larger sky coverage. In order to successfully extract all cosmological and astrophysical information from large-scale structure, these developments must be accompanied by an improved theoretical understanding of the statistical properties of galaxy populations on cosmological scales. 

In stark contrast to the (primary) anisotropies of the cosmic microwave background, the distribution of LSS is highly non-Gaussian due to nonlinear gravitational evolution of the initial density perturbations. Hence, the statistics of LSS are not captured by only the two-point correlation function and linear theory does not suffice except at very large separations. This has triggered a surge in the study of statistics of LSS beyond the power spectrum, the simplest of which is the bispectrum, i.e. the Fourier transform of the three-point correlation function \cite{sefusatti_bisp}. In the case of correlations between galaxy positions, the addition of the bispectrum has been shown to yield a gain in constraining power on the cosmological model parameters.  \cite{damico_oneloopbisp, philcox_bisp,ivanov_bisp,cabass1,cabass2}.

However, in modern surveys not only the positions of galaxies, but also their \textit{shapes} can be inferred (where reliable imaging is available). Correlations of shapes of galaxies are a rich source of cosmological information, both as tracers of weak gravitational lensing and through \textit{intrinsic alignment} (IA) (see e.g. \cite{troxel_ishak} for a review). Weak lensing affects the apparent orientations of galaxies due to light bending by the intervening matter distribution along the line of sight. By contrast, IA arises due to the correlation of shapes of galaxies with the local gravitational tidal field \cite{catelan_la,hirata_seljak}. Similarly to galaxy positions, the statistics of shapes of galaxies are thus also non-Gaussian. For that reason, improving our understanding of non-Gaussian statistics of shape fields is a natural avenue to achieve better modelling of LSS.  

The modeling of higher-order statistics of weak gravitational lensing has been considered in numerous other works (see e.g. \cite{shear_3pt1,shear_3pt2,shear_3pt3,burger} for recent applications of three-point statistics, or \cite{howls1, howls2} for more general higher-order statistics). However, the contribution of IA to the observed three-point signal has received less attention (though see \cite{pyne_3pt,linke_lowz}) than the two-point IA signal, which has been extensively studied in e.g. \cite{kurita_ia,okumura_nl,taruya_rsd}. It is of note that this contribution may contain valuable information on interesting physical phenomena such as anisotropic local primordial non-Gaussianity \cite{akitsu_png,kurita_png}, in analogy to the galaxy bispectrum. 

The objective of this paper and its earlier companion \cite{bakx_bisp} (hereafter Paper I) is to assess the gain in information on shape statistics by including the IA bispectrum. On the modelling side, we restrict ourselves to large scales where the bispectrum signal can be reliably computed using perturbation theory. The general perturbative framework developed in \cite{vlah_eft1,vlah_eft2,bakx_eft} (see also \cite{chen_lpt}) offers a systematic improvement over simpler models of IA, such as the `NLA' bispectrum model used in \cite{pyne_3pt}. As we will see in this work, the higher-order corrections to the IA bispectrum are clearly present for the halo sample we consider and cannot be neglected. In addition, we will assume that three-dimensional positions of galaxies are available through spectroscopic redshift measurements. In that case, the bispectrum of intrinsic alignment can be summarized in terms of multipole moments with respect to the line of sight, for which Paper I developed separable estimators. These estimators capture the full scale dependence and angular structure of the bispectrum and do not restrict to specific configurations such as equilateral triangles or triangles in the sky plane. Notably, such estimators are applicable regardless of the scales being modelled (as long as a flat-sky treatment is appropriate). In Paper I, we forecasted an SNR of $\approx 30$ for $B_{DDE}$ (where $D$ denotes the galaxy density) in a DESI-like LRG sample, where shapes are measured using LSST. Direct measurements of intrinsic alignments for samples with spectroscopic redshifts can aid in isolating this effect from cosmic shear \cite{samuroff_ia,hervaspeters}. 

In this work, we apply the formalism developed in Paper I to measure multipoles of alignment bispectra of halos in N-body simulations. This serves as a vital cross-check for our theory model as well as a test bed for the implementation of the estimators from Paper I. Since our focus here is on the consistency with the halo alignment - dark matter power spectrum from \cite{bakx_eft}, we restrict ourselves to cross-bispectra of halo shapes with the dark matter density field. In Section \ref{sec:theory} we provide a brief summary of the underlying theory model based on perturbation theory, as well as the flat-sky projection and the definition of the bispectrum multipoles and refer the reader to Paper I for more details. Then Section \ref{sec:sims} summarizes the specifics of the simulation suite used and our methodology for analysing the bispectra. Section \ref{sec:results} summarizes the results we obtained as well as various cross-checks of approximations for the covariance and data vector calculation. We then conclude in Section \ref{sec:discussion}. Some explicit formulae for the theory model are provided in Appendix \ref{sec:expr}. Appendices \ref{sec:cov} and \ref{sec:los} provide further details on the comparison between the Gaussian covariance and the sample covariance of the IA bispectrum as well as potential line-of-sight dependencies of the projected bispectra.

Our Fourier conventions are
\begin{equation}
\begin{aligned}
    f(\bb{k}) &= \int \md^3 \bb{x}\, f(\bb{x})e^{\mi \bb{k}\cdot \bb{x}}; \nonumber \\ 
    f(\bb{x}) &= \int \frac{\md^3 \bb{k}}{(2\pi)^3} \,e^{-\mi \bb{k}\cdot \bb{x}}f(\bb{k}).
    \end{aligned}
\end{equation}
We use $c=1$ throughout. The cosmology used for the theory computation and simulations is Euclidean $\Lambda$CDM with $\omega_b=0.02225, \omega_c=0.1198,\Omega_\Lambda = 0.6844, n_s=0.9645,\ln(10^{10}A_s)=3.094$ which is consistent with the {\it Planck} data \cite{Planck2018}. The linear power spectrum was computed using \texttt{CAMB} \cite{camb}. 
\section{Tree-level Shape Bispectrum}\label{sec:theory}
The tree-level tensor bispectrum in the Effective Field Theory for Intrinsic Alignments (EFT of IA) was analyzed in Paper I. For this reason, we refer the reader to this paper for all the relevant details and restrict ourselves to a summary of its results (see also \cite{vlah_eft1,vlah_eft2,bakx_eft} for more details). 

To describe the shape field $g_{ij}$, which is a traceless 2-tensor defined from the inertia tensor of the halo $I_{ij}$ at position $\bb{x}$ as 
\begin{equation}
    g_{ij}(\bb{x}) = \frac{\text{TF}(I_{ij})({\bf x})}{\langle\text{Tr}(I_{ij})\rangle},
\end{equation}
one needs the following operator expansions: 
\begin{equation}
\begin{aligned}
    g_{ij}(\bb{x}) &=  g_{ij}^{\text{loc}}(\bb{x}) + g_{ij}^{\text{stoch}}(\bb{x}); \\
    g_{ij}^{\text{loc}}(\bb{x}) &= \sum_\mathcal{O} b^\text{g}_\mathcal{O} \text{TF}(\mathcal{O}_{ij})(\bb{x}); \\
    g_{ij}^{\text{stoch}}(\bb{x}) &= \sum_\mathcal{O} \epsilon_\mathcal{O}(\bb{x}) \text{TF}(\mathcal{O}_{ij})(\bb{x}). 
\end{aligned}
\end{equation}
The deterministic part, consisting of combinations of operators $\Pi_{ij}^{[n]}$ with $n=1,2$ \cite{mirbabayi}, contains four operator biases $b_1^\text{g}, b_{2,1}^\text{g},b_{2,2}^\text{g},b_{2,3}^\text{g}$. The stochastic fields possess non-trivial N-point functions for all N $>1$ and have vanishing mean \cite{schmidt_bias}. Some examples of the contributions arising from these fields are 
\begin{equation}
    g_{ij}^{\text{stoch}}(\bb{x}) \supset \{\epsilon_{ij}(\bb{x}); \epsilon^\delta_{ij}(\bb{x})\delta(\bb{x});\epsilon^\Pi(\bb{x})\text{TF}(\Pi_{ij}^{[1]}(\bb{x})), \dots  \}.
\end{equation}
As in Paper I, we did not consider higher derivative contributions to $g_{ij}$ (or the matter density field). Assuming that the non-locality scale of halos is similar to (or smaller than) the nonlinear scale, such contributions would be of higher order than the ones we include in the tree-level bispectrum. We also did not consider correlations with scalar biased tracers in this paper. 
\subsection{Tree-level Bispectrum}
Introducing the shorthand 
\begin{equation}
    S_{ij}(\bb{x}) = \frac{1}{3}\delta_{ij}^\mathrm{K}\delta(\bb{x}) + g_{ij}(\bb{x}),
\end{equation}
for the combination of the halo shape field and the dark matter density, the full bispectrum $B_{ijklrs}$ we consider in this work is
\begin{equation}\label{eq:bispdef}
\begin{aligned}
     \langle S_{ij}(\mathbf{k}_1)S_{kl}(\mathbf{k}_2)S_{rs}(\mathbf{k}_3) \rangle &= (2\pi)^3 \delta^D(\mathbf{k}_1+\mathbf{k}_2+\mathbf{k}_3) \\
     &\times B_{ijklrs}(\mathbf{k}_1,\mathbf{k}_2,\mathbf{k}_3).
\end{aligned}
\end{equation}
Here $\delta(\bb{x})$ is the matter overdensity and $\delta_{ij}^\mathrm{K}$ is the Kronecker delta. We then obtain the mixed scalar ($\delta$) and tensor (g) bispectra by taking trace and trace-free parts of the full tensorial bispectrum: 
\begin{equation}\label{eq:tensbisp}
\begin{aligned}
    B^{\delta\delta\text{g}}_{ij} &= \langle \text{Tr}(S)(\mathbf{k}_1)\text{Tr}(S)(\mathbf{k}_2)\text{TF}(S_{ij})(\mathbf{k}_3) \rangle ' \\
    &= \langle \delta(\mathbf{k}_1)\delta(\mathbf{k}_2)g_{ij}(\mathbf{k}_3) \rangle', \\
    B^{\delta\text{gg}}_{ijkl} &= \langle \text{Tr}(S)(\mathbf{k}_1)\text{TF}(S_{ij})(\mathbf{k}_2)\text{TF}(S_{kl})(\mathbf{k}_3) \rangle ' \\
    &= \langle \delta(\mathbf{k}_1)g_{ij}(\mathbf{k}_2)g_{kl}(\mathbf{k}_3) \rangle', \\
    B^{\text{ggg}}_{ijklrs} &= \langle \text{TF}(S_{ij})(\mathbf{k}_1)\text{TF}(S_{kl})(\mathbf{k}_2)\text{TF}(S_{rs})(\mathbf{k}_3) \rangle ' \\
    &= \langle g_{ij}(\mathbf{k}_1)g_{kl}(\mathbf{k}_2)g_{rs}(\mathbf{k}_3) \rangle'.
\end{aligned}   
\end{equation}
We used the prime to simplify the notation by omitting the Dirac delta that enforces momentum conservation. In this work, we will only focus on the leading order bispectrum in perturbation theory. It consists of a deterministic part (involving only the operator biases associated to the $\Pi_{ij}^{[n]}$) and a stochastic part involving one or more terms from the expansion of $g_{ij}^\text{stoch}$:
\begin{equation}
    B_{ijklrs}^\text{tree} = B_{ijklrs}^{\text{det,tree}} + B_{ijklrs}^{\text{stoch,tree}}. 
\end{equation}
Ignoring leading stochastic terms $\sim \langle \epsilon \epsilon \epsilon \rangle $ (which, as we will see later, indeed do not contribute to any of the bispectra listed in \eqref{eq:tensbisp} after projection onto the sky) we can write all bispectrum contributions as products of one second order field and two first order fields (indicated by superscripts), viz. 
\begin{equation}
    \begin{aligned}
        B_{ijklrs}^{\text{tree}}(\mathbf{k}_1,\mathbf{k}_2,\mathbf{k}_3) = \langle & S_{ij}^{(2)}(\mathbf{k}_1)S_{kl}^{(1)}(\mathbf{k}_2)S_{rs}^{(1)}(\mathbf{k}_3) \rangle' \\
   &+ \text{2 perm.} 
    \end{aligned}
\end{equation} 
Here the second order field may refer to either a second order operator bias contribution or a (Fourier transformed) stochastic contribution of the form $\propto \epsilon(\bb{x})\delta(\bb{x})$. All the relevant contributions (including the stochastic terms) are listed in Appendix \ref{sec:expr}. Since we only consider correlations of the shape field with the matter overdensity, we do not consider any of the scalar bias parameters $b_1^\text{s}, b_{2,1}^\text{s}, b_{2,2}^\text{s} $ from Paper I. Furthermore, since the dark matter density field does not contain any stochastic contribution at this order, the only stochastic bias parameter left (after projecting bispectra onto E- and B-modes as in the next Section) is $\alpha_\text{gg}^{1\delta}$. It occurs in all bispectra with more than one shape field. Thus, the bispectra from Eq. \eqref{eq:tensbisp} depend on five parameters in total, namely  
\begin{equation}\label{eq:modelpars}
\{b_1^\text{g},b_{2,1}^\text{g}, b_{2,2}^\text{g},b_{2,3}^\text{g},\alpha_\text{gg}^{1\delta}\}.
\end{equation} 
The three second order IA parameters are simply a redefinition of the more commonly used `TATT' parameters $b_{KK}, b_t, b_{\delta K}$ \cite{blazek_tatt,bakx_eft}. It turns out that second order IA contributions are generally smaller than the linear alignment term proportional to (powers of) the linear shape bias \footnote{Indeed, \cite{linke_lowz} find that a simpler alignment model is sufficient to describe three-point alignments in the BOSS LOWZ sample.}. However, as we will see, the halo sample considered here cannot be modelled without second order IA contributions. The leading stochastic shape bispectrum $\langle \epsilon_{ij} \epsilon_{kl} \epsilon_{rs} \rangle \propto \alpha_{\text{ggg}}^{111}$ turns out to vanish after projecting the tensor bispectrum onto the sky (see App. A of Paper I), which is the part we discuss next.
\subsection{Flat Sky Projection} 
Not all five degrees of freedom of the shape field can be discerned by an observer at a fixed location. Instead, only two degrees of freedom in $g_{ij}$ are observable. They constitute a spin-2 field on the sky which is commonly decomposed into E- and B-modes in Fourier space. Throughout this work we will use the flat-sky approximation to construct projected observables. For an in-depth study of the validity of the flat-sky approximation for various tracers, see e.g. \cite{gao}. Note that current surveys may require wide-angle effects to be taken into account \cite{philcox_re} - we leave this for future work. To recap the findings of \cite{vlah_eft1,vlah_eft2}, the projection on the sky is performed by applying the trace-free operator 
\begin{equation}
     \mathcal{P}^{ijkl}(\hat{\bb{n}}) = \text{TF}(\mathcal{P}^{ik}(\hat{\bb{n}})\mathcal{P}^{jl}(\hat{\bb{n}}))
\end{equation}
to the 3D shape field $g_{ij}$, thus yielding 
\begin{equation}
    \gamma_{ij,I}(\mathbf{x}) = \mathcal{P}^{ijkl}(\hat{\bb{n}}) g_{kl}(\bb{x}).
\end{equation}
The two remaining degrees of freedom are then 
\begin{equation}
    \gamma_{\pm 2}(\bb{x}) := (\bb{M}^{\pm 2}_{ij})^*(\hat{\bb{n}})\gamma_{ij,I}(\mathbf{x}) 
\end{equation}
with 
\begin{equation}
\begin{aligned}
    \mathbf{M}_{ij} ^{(\pm 2)}(\hat{\bb{n}}) &:= \mathbf{m}_i ^\pm \mathbf{m}_j ^\pm;
\end{aligned}
\end{equation}
where $\mathbf{m}_1 = \hat{\bb{x}}$, $\mathbf{m}_2 = \hat{\bb{y}}$ and $\mathbf{m}^\pm := \mp \frac{1}{\sqrt{2}}(\mathbf{m}_1 \mp \mi\mathbf{m}_2)$. Finally, the E- and B-modes are
\begin{equation}
\begin{aligned}
    \gamma_E(\bb{k}) &= \frac{1}{2}(\gamma_{+2}\exp (-2\mi\phi_{\bb{k}})+\gamma_{-2}\exp (+2\mi\phi_{\bb{k}})); \\
    \gamma_B(\bb{k}) &= \frac{1}{2\mi}(\gamma_{+2}\exp (-2\mi\phi_{\bb{k}})-\gamma_{-2}\exp (+2\mi\phi_{\bb{k}}));
\end{aligned} 
\end{equation}
where $\phi_{\bb{k}}$ is the polar angle of the wavevector in the plane perpendicular to $\bb{\hat{n}}$, i.e. 
\begin{equation}
    \exp (+\mi\phi_{\bb{k}}) = \frac{\bb{k}_x + \mi \bb{k}_y}{\sqrt{\bb{k}_x^2+\bb{k}_y^2}}.
\end{equation}
We will instead work with projectors onto E- and B-modes
\begin{equation}
    \begin{aligned}
        \bb{M}^E_{ij}(\hat{\bb{k}},\hat{\bb{n}}) = \frac{1}{2}\big( & \bb{M}^{(-2)}_{ij}(\hat{\bb{n}})\exp (-2\mi\phi_{\bb{k}}) \\
        &+\bb{M}^{(+2)}_{ij}(\hat{\bb{n}})\exp (+2\mi\phi_{\bb{k}})\big) 
    \end{aligned}
\end{equation}
and
\begin{equation}
    \begin{aligned}
        \bb{M}^B_{ij}(\hat{\bb{k}}, \hat{\bb{n}}) = \frac{1}{2\mi}\big( &\bb{M}^{(-2)}_{ij}(\hat{\bb{n}})\exp (-2\mi\phi_{\bb{k}}) \\
        &-\bb{M}^{(+2)}_{ij}(\hat{\bb{n}})\exp (+2\mi\phi_{\bb{k}})\big).
    \end{aligned}
\end{equation}
The E- and B-modes can then be obtained by applying this projector to $g_{ij}(\bb{x})$, since it commutes with the flat-sky projection operator $\mathcal{P}_{ijkl}$ \cite{vlah_eft1,bakx_eft}: 
\begin{equation}
\begin{aligned}
    \gamma_E(\bb{k},\hat{\bb{n}}) &= \bb{M}^E_{ij}(\hat{\bb{k}},\hat{\bb{n}})^*g_{ij}(\bb{k});\\ 
    \gamma_B(\bb{k},\hat{\bb{n}}) &= \bb{M}^B_{ij}(\hat{\bb{k}},\hat{\bb{n}})^*g_{ij}(\bb{k}).
\end{aligned}
\end{equation}
Introducing the notation 
\begin{equation}
    \bb{M}_{ij}^\delta(\hat{\bb{k}}) = \delta_{ij}^\mathrm{K}
\end{equation}
and dropping the dependence of the projectors on $\hat{\bb{n}}$, we can form the fields 
\begin{equation}
    X(\bb{k}) = \bb{M}_{ij}^X(\hat{\bb{k}})S_{ij}(\bb{k})
\end{equation}
for $X=\delta,E,B$, and the corresponding bispectra 
\begin{equation}\label{eq:indep}
    B_{XYZ} = \bb{M}_{ij}^{X}(\hat{\bb{k}}_1) \bb{M}_{kl}^{Y}(\hat{\bb{k}}_2)\bb{M}_{rs}^{Z}(\hat{\bb{k}}_3)B_{ijklrs}. 
\end{equation}
Thus, the projection of the shape field onto the sky gives rise to the following bispectra from $S_{ij}$: 
\begin{equation}\label{eq:combs}
    \begin{aligned}
        \delta\delta\delta&: B_{\delta\delta\delta}; \\
        \delta\delta\text{g}&: B_{\delta\delta E}, B_{\delta\delta B}; \\
        \delta\text{gg}&: B_{\delta EE}, B_{\delta EB}, B_{\delta BB}; \\
        \text{ggg}&: B_{EEE}, B_{EEB}, B_{EBB}, B_{BBB}. \\
    \end{aligned}
\end{equation}
We emphasize that all these combinations are nonzero even in the absence of parity violation; the vanishing of power spectra containing a single B-mode does not generalize to higher-order statistics such as the bispectrum (see again Paper I for more details). 

Since the linear order B-field vanishes, i.e. 
\begin{equation}\label{eq:bmodezero}
    B^{(1)}(\bb{k}) = \bb{M}_{ij}^{B}(\hat{\bb{k}})S_{ij}^{(1)}(\bb{k}) = 0,
\end{equation}
the spectra with two B-modes (that is, $B_{\delta BB}$ and $B_{EBB}$) necessarily only depend on the stochastic parameter $\alpha_\text{gg}^{1\delta}$; they do not depend on the $b_{2,a}^\text{g}$. The bispectrum consisting of three B-modes, $B_{BBB}$ vanishes identically in our model since the contribution from $\alpha_\text{gg}^{1\delta}$ also gets projected out, i.e. it vanishes when applying the B-mode projector \cite{bakx_bisp}. Furthermore, the deterministic piece of spectra containing any B-modes is typically strongly suppressed, and as such we do not expect these spectra to meaningfully constrain the $b_{2,a}^\text{g}$ bias parameters. For that reason, we will not consider the parity-odd bispectra $B_{\delta \delta B}$ and $B_{\delta E B}$ in any baseline analysis; instead we will only focus on fitting parity-even bispectra and subsequently check that the obtained results are broadly consistent with the parity-odd sector.

A note of caution: in \cite{bakx_eft} (see their App. B) it was found that the definition of E- and B-modes as usually employed in cosmological simulations is in principle different from the way these projected quantities are treated in the theoretical modelling. Concretely, the theoretical model for the E- and B-modes should receive corrections at second and higher orders in perturbation theory that depend on the line of sight, similar to the case of selection effects for galaxy clustering \cite{agarwal}. At face value, since the bispectrum involves second order fields, these corrections should be taken into account here. However, such contributions turn out to be proportional to squares of shape bias parameters \cite{bakx_eft} and are thus very suppressed (see Appendix \ref{sec:los} for an assessment of their importance). We will from now on ignore this issue in the main text. 

\subsection{Bispectrum Multipoles}
The bispectrum of tensor fields admits a multipole decomposition as detailed in Paper I. There it was argued that a separable local plane-parallel Yamamoto estimator requires the use of \textit{associated} Legendre polynomials to define the multipole moments \cite{kurita_window, inoue}. In that work, it was also shown that differences between the use of ordinary and associated Legendre polynomials are small and that associated Legendre polynomials appear better suited for $B_{\delta\delta E}$ (at least in real space, which is the case we restrict to here). For that reason, we will only consider associated Legendre multipoles (ALMs) here. We thus also omit the superscript `A' on the multipoles that was present in Paper I to distinguish between ordinary and associated Legendre multipoles, and simply refer to the ALMs as multipoles. Specifically, defining the even and odd projectors
\begin{equation}\label{eq:asproject}
\begin{aligned}
     \mathcal{P}_{XYZ}^{\ell_1\ell_2}(\{\hat{\bb{k}}_i\})_\text{even} &= 
     \mathcal{L}^{m_X}_{\ell_1}(\mu_1)\mathcal{L}^{m_Y}_{\ell_2}(\mu_2)\mathcal{L}^{m_Z}_{m_Z}(\mu_3);\\
     \mathcal{P}_{XYZ}^{\ell_1\ell_2}(\{\hat{\bb{k}}_i\})_\text{odd} &=
     \mathcal{L}^{m_X}_{\ell_1}(\mu_1)\mathcal{L}^{m_Y}_{\ell_2}(\mu_2)\mathcal{L}^{m_Z}_{m_Z}(\mu_3)\\
     &\times \big(\hat{\bb{n}} \cdot (\hat{\bb{k}}_1 \times \hat{\bb{k}}_2)\big),
\end{aligned}
\end{equation}
the multipoles are given as 
\begin{equation}\label{eq:olmdef}
\begin{aligned}
    B^{\ell_1\ell_2}_{XYZ}(k_1,k_2,k_3) &= \frac{N_{\ell_1\ell_2}}{4\pi}\int\md\mu_1 \md\xi \bigg(\mathcal{P}_{XYZ}^{\ell_1\ell_2}(\{\hat{\bb{k}}_i\}) \\
    &\times B_{XYZ}(k_1,k_2,k_3,\mu_1,\xi) \bigg) 
\end{aligned}
\end{equation}
with the normalization constant
\begin{equation}\label{eq:nml}
    N_{\ell_1\ell_2} = (2\ell_1+1)(2\ell_2+1).
\end{equation}
Here the integration is over the angles $\mu_1$ and $\xi$ as first introduced in \cite{scoccimarro_fast}, which describe the orientation of the bispectrum with respect to the line of sight. This orientation requires only two angles to be specified, because statistical isotropy demands that there is no dependence on any azimuthal rotations around the line of sight.
The $\mu_i$-dependence is captured via the associated Legendre polynomials $\mathcal{L}^{m}_{\ell}(\mu_i)$. We only need two independent multipole indices to capture all angular dependence, since the third angle $\mu_3$ is always linearly dependent on the other two. In Paper I, we also showed in App. B that a double expansion in powers of $\mu_1$ and $\mu_2$ is equivalent to the one in terms of spherical harmonics $Y_{\ell m}(\mu_1,\xi)$.
\section{Simulations and Data Analysis}\label{sec:sims}
We will use a suite of 20 high-resolution dark-matter only N-body simulations from the \texttt{DarkQuest} Project \citep{dark_quest}. Each realization has $2048^3$ 
N-body particles with mass $M_p=1.02 \times 10^{10} M_{\odot}/h$ and box side length $L=1 \text{ Gpc}/h$. For more specifics, see e.g. \cite{kurita_ia}. In this setup, we will take the halo sample in mass range $10^{12}-10^{12.5} M_{\odot}/h$ at redshift $z=0$, which we also analyzed in \cite{bakx_eft}.  This is especially convenient since it allows us to check whether bias parameters that are shared between the power spectrum and bispectrum of IA match. We defer performing a joint analysis to potential future work. The sample of halos in a single realization has different properties from the Luminous Red Galaxy (LRG) sample considered in Paper I. Notably, the number density is roughly $\bar{n} = 3\times 10^{-3}(h/\text{Mpc})^3$ and $A_\text{IA} \approx 8$ compared to $\bar{n} = 5\times 10^{-4}(h/\text{Mpc})^3$ and the alignment amplitude (defined in Eq. \eqref{eq:aiadef}) $A_\text{IA} \approx 4$ in Paper I, meaning that the signal-to-noise ratio (SNR) for a given total volume is much higher. This allows us to perform a more stringent test of the model under consideration. Higher alignment amplitudes of $A_\text{IA} \sim 15$ arise for galaxy clusters, although such samples would have much lower number density than LRGs \cite{shi_clusters}.
\subsection{Estimator}
In Paper I we constructed separable estimators for the bispectrum multipoles, which can be computed using FFTs. They are given by
\begin{equation}\label{eq:est}
\begin{aligned}
    \hat{B}^{\ell_1\ell_2}_{XYZ}(& k_1,k_2,k_3) = \frac{N_{\ell_1\ell_2}}{N_\Delta V}\sum_{\mathbf{q}_i \in \text{bin } i}\delta^K(\mathbf{q}_{123}) \\
    &\times \hat{X}(\mathbf{q}_1) \hat{Y}(\mathbf{q}_2)\hat{Z} (\mathbf{q}_3)\mathcal{P}_{XYZ}^{\ell_1\ell_2}(\{\hat{\bb{q}}_i\}),
\end{aligned}
\end{equation}
where $N_\Delta$ is the number of closed triangles in the bin. The estimators for the bispectra are given by explicitly summing all wavevector triplets $(\mathbf{k}_1,\mathbf{k}_2,\mathbf{k}_3)$ that enter a given triangle bin. Importantly, we do not employ any shot noise subtraction schemes for the bispectrum.  Furthermore, the results of Paper I also justify symmetrizing the \textit{parity-even} multipoles over all six triangle side permutations. We thus consider the symmetrized estimator
\begin{equation}\label{eq:altdef}
    \widetilde{B}_{XYZ}^{\ell_1\ell_2}(k_1,k_2,k_3) =\frac{1}{6}\sum_{\sigma \in S_3} B^{\ell_1\ell_2}_{XYZ}(k_{\sigma(1)},k_{\sigma(2)},k_{\sigma(3)})
\end{equation}
when $XYZ$ is parity-even. This renders the covariance matrix diagonal in triangle space in the Gaussian approximation \cite{bakx_bisp}. To this end, for the parity-even sector we will consider only ordered triangle bins which satisfy $k_1 \geq k_2 \geq k_3$. When enumerating these bins, they are ordered `lexicographically', i.e. the first few bins are $(1,1,1),(2,1,1),(2,2,1),(2,2,2), (3,1,1), \dots $ et cetera. Note that this setup includes open triangles such as $(2,1,1)$ for which the bin centers do not form a closed triangle; we discuss this in more detail below. We use a radial bin width of $\Delta k =0.4/30 \,h$/Mpc which approximately equals $2k_f$, where $k_f = 2\pi / L$ is the fundamental mode. The bin width is chosen to strike a balance between having too few modes per bin (and hence very large discreteness effects) and loss of information due to over-compressing the data vector.  For example, for triangles with side lengths up to the maximum wavenumber $k_\text{max} = 0.2\,h/\text{Mpc}$ there are $15$ radial wavenumber bins, resulting in a total of $428$ ordered triangle bins. For the parity-odd bispectra, which have an odd number of B-modes, i.e. $B_{\delta \delta B}, B_{\delta E B}, B_{EEB}$ and $B_{BBB}$ we will consider all triangles with the sole restriction $k_1 > k_2$; this results in $917$ triangles in the parity-odd sector for $k_\text{max} = 0.2\,h/\text{Mpc}$. Note that this restriction must necessarily be applied for all bispectra except for $B_{\delta EB}$, since they are symmetric in their first two arguments. In the case of $B_{\delta EB}$, we add to these triangles also the ones with $k_1 = k_2 = k_3$, of which there are 15. Under these assumptions, the Gaussian covariance of all multipoles (both even and odd) is diagonal in the triangle bins\footnote{In the even case, this was already argued in Paper I. For the odd case, it can be shown from Eq. (E9) from Paper I by observing that $P_{XB}(\bb{k}) = 0$ whenever $X \neq B$. The parity-odd covariance would not be diagonal if we also included the $k_1 > k_2$ part of the data vector for $B_{\delta EB}$.} and also has no cross-correlation between even and odd multipoles. 

We will not include parity-odd bispectra in the likelihood at any point, since they are detected at small significance compared to the parity-even ones in a single simulation volume. However, we emphasize that these spectra are clearly nonzero and as we will show, their amplitudes are fully consistent with what is inferred from fitting the parity-even bispectra.  
\subsection{Covariance}
We will explore two choices for evaluating the bispectrum covariance, namely (i) the measured sample variance from simulations \textit{where all elements that are zero in the Gaussian approximation are set to zero}, or (ii) the Gaussian covariance from Paper I with the NLA model inserted for the power spectra:
\begin{equation}\label{eq:nlamod}
\begin{aligned}
    P_{EE}(k,\mu) &= \frac{1}{4}(b_1^{\text{g}})^2(1-\mu^2)^2P_{\text{NL}}(k) + c^{\text{g}}; \\
    P_{\delta E}(k,\mu) &= \frac{1}{2}b_1^{\text{g}}(1-\mu^2)P_{\text{NL}}(k); \\
    P_{BB}(k,\mu) &= c^{\text{g}}; \\
    P_{\delta\delta}(k) &= P_{\text{NL}}(k), 
\end{aligned}
\end{equation}
where $c^{\text{g}}$ is the constant shape noise term\footnote{This parameter was labeled $\alpha_\text{gg}^{11}$ in Paper I.}.
Here the dimensionless bias parameter $b_1^\text{g}$ is related to the commonly used $A_{\text{IA}}$ parameter via
\begin{equation}\label{eq:aiadef}
    b_1^{\text{g}} = -2A_\text{IA} C_1 \rho_\text{cr} \Omega_{m,0} / D(z)
\end{equation}
and $P_\text{NL}$ is computed using the \texttt{CCL} wrapper for \texttt{HaloFit} \cite{ccl,halofit}. In the expression for $b_1^\text{g}$, $D(z)$ is the linear growth factor normalized to $1$ today and $C_1\rho_\text{cr} = 0.0134$ \cite{hirata_seljak,bridle_king}. For the Gaussian covariance, we use the fiducial set of values 
\begin{equation}
\begin{aligned}
    b_1^\text{g} &= -0.078; \\
    c^\text{g} &= 9.6;
\end{aligned}
\end{equation}
consistent with \cite{bakx_eft}. The expression for the Gaussian covariance is rather cumbersome and we do not reproduce it here, but rather refer to App. E of Paper I. The Gaussian covariance assumption predicts that correlations between bispectrum multipoles in different triangle bins are zero. 

Due to the limited number of realizations available to us, the sample covariance for the bispectrum is very noisy\footnote{Observe that generating many independent N-body simulations to measure shapes is especially challenging; the mass resolution needs to be sufficiently high in order to reliably resolve halo shapes from their inertia tensor \cite{herle}.}. Thus, we are unable to probe cross-covariances between different triangle shapes, which are expected to be subdominant. However, at least for identical triangle configurations we can perform a meaningful comparison between the Gaussian approximation and the measured variance of a multipole in a given bin. We can do the same for cross-covariances between different multipoles in the same triangle bin. Formally, for two arbitrary bispectrum multipoles $B_L,B_{L'}$ and triangles $T,T'$ we consider the `block diagonal sample variance' estimator\footnote{Moreover, when $T=T'$ we also set all elements to zero that are expected to be zero in the Gaussian approximation. These are elements where a bispectrum without any B-modes is correlated with a bispectrum that does contain B-modes, e.g. $B^{00}_{\delta \delta E}$ and $B_{\delta BB}^{02}$. This cross-correlation is zero in the Gaussian approximation since $P_{XB}(\bb{k}) = 0$ whenever $X \neq B$.} 
\begin{equation}\label{eq:covest}
    \hat{C}_{LL'}(T,T') = \frac{\delta_{TT'}}{N_r-1}\sum_r(\hat{B}_{r,L}^T-\hat{\bar{B}}^T_L)(\hat{B}^{T'}_{r,L'}-\hat{\bar{B}}^{T'}_{L'}),
\end{equation}
\begin{figure}[t]
    \centering
    \includegraphics[width=\linewidth]{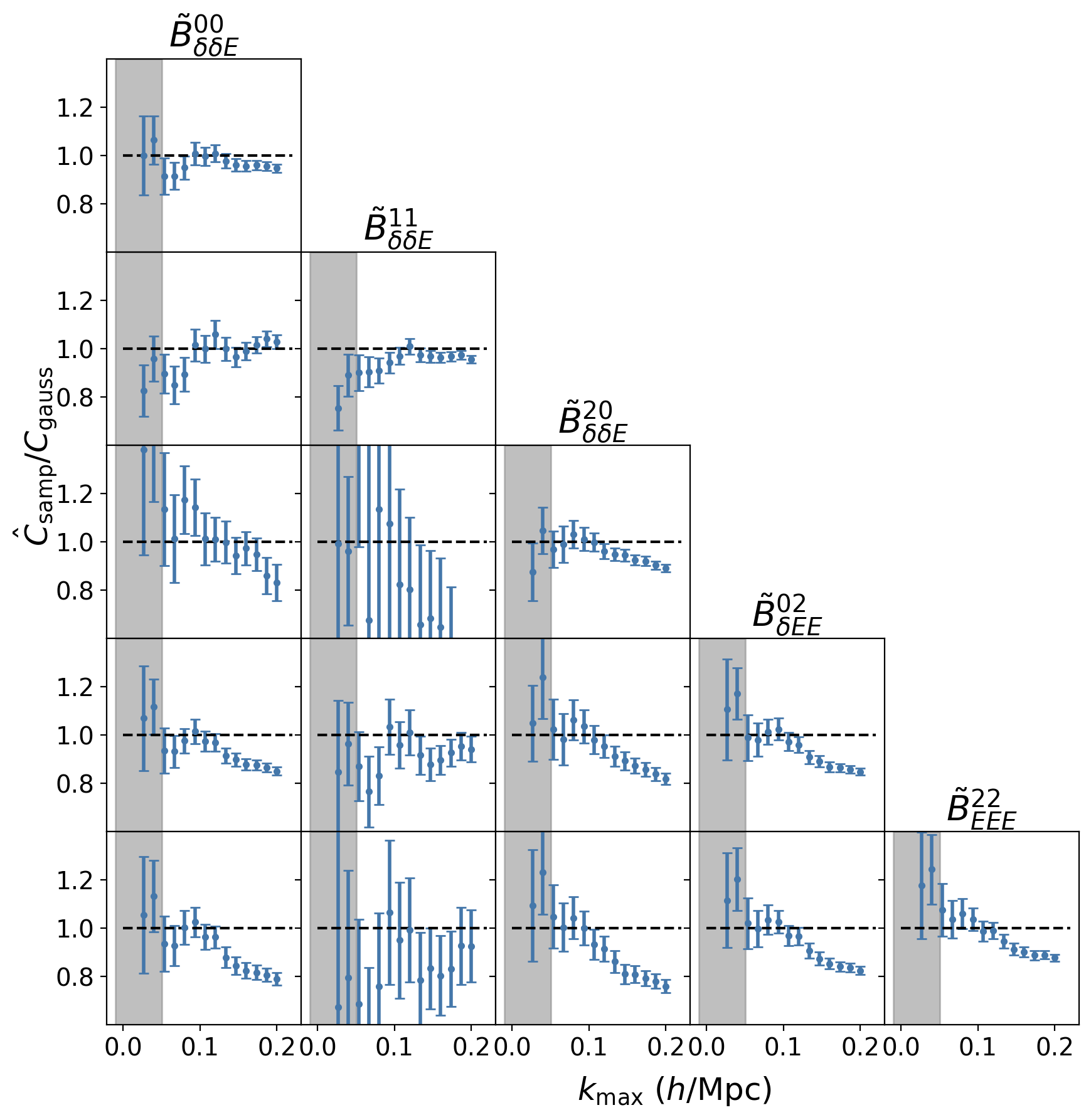}
    \caption{Ratio of diagonal sample covariance to the theoretical Gaussian prediction for several different multipoles. The grey shade indicates the region where $k_\text{max} < 0.05\,h/\text{Mpc}$ (this region was chosen somewhat artifically); deviations in this region are likely due to discretization effects and do not indicate model failure. }
    \label{fig:cov_comp}
\end{figure}
where the sum is over all $N_r = 20$ realizations, $\hat{B}_{r,L}^T$ is the measurement of multipole $L$ in triangle $T$ in the realization $r$ and $\hat{\bar{B}}_{L}^T$ denotes the sample mean of $\hat{B}^T_{r,L}$. The `block' thus refers to a block \textit{in multipole space} where the covariance has off-diagonal elements (in the $L$ and $L'$ indices), since different multipoles have non-trivial cross-covariance for the same triangle even in the Gaussian approximation. Setting the off-diagonal elements \textit{in triangle space} (in the $T$ and $T'$ indices) to zero through $\delta_{TT'}$ in the estimator effectively reduces the impact of the noise when inverting the matrix\footnote{In fact, the estimator would be singular if we did not do this, since we would have more triangles than realizations.}. 

The comparison of Eq. \eqref{eq:covest} with the theoretical Gaussian covariance estimate is shown in Figure \ref{fig:cov_comp} for five different multipoles involving the $\delta$ and $E$ fields. For several increasing values of $k_{\text{max}}$, we consider the ratio of the sample and Gaussian covariances for all triangles with all sides below $k_{\text{max}}$ and plot the mean and sample variance of these ratios to reduce scatter. On large scales ($k<0.10\,h$/Mpc), the Gaussian approximation works well (aside from very large scales $k<0.05\,h$/Mpc, where discreteness effects occur in the covariance matrix), while for larger $k_\text{max}$ the Gaussian covariance (using the nonlinear matter power spectrum) tends to somewhat overpredict the error. 
Moreover, we observe at most $\sim 20\%$ offsets between the theoretical, Gaussian covariance and Eq. \eqref{eq:covest}, which amounts to $\sim 10\%$ on the corresponding errors. We consider this sufficient for the purpose of this work. We do not use any large-scale cuts (i.e. $k_\text{min}$) in the remainder of the paper, since large scales do not impact constraints significantly.

Supposing that the true covariance is diagonal in triangle space (as in the Gaussian approximation), the estimator in Eq. \eqref{eq:covest} is unbiased. Its inverse can be computed by simply inverting, for every triangle bin $T$, the $N_L\times N_L$ block in multipole space, where $N_L$ is the number of multipoles. The resulting inverse is however not unbiased and needs to be `Hartlap-corrected' \cite{hartlap}. This correction needs to applied to each of the blocks, since the expectation of the inverse of the estimator is given by \cite{anderson2003}
\begin{equation}\label{eq:hartlapfac}
    \langle(\hat{C}_{LL'})^{-1}\rangle = \frac{N_r-1}{N_r-N_L-2} (C_{LL'}^{\text{true}})^{-1}.
\end{equation}
which amounts to an increase as large as $\sim 50\%$ when e.g. $N_r = 20$ and $N_L = 5$ (and thus roughly an increase of $\sim 25\%$ in the posteriors).

\subsection{Discreteness and Binning Effects}
It is customary to replace the discrete sum over wavevectors in the estimator \eqref{eq:est} by a continuous integration, i.e. 
\begin{equation}\label{eq:intapprox}
\begin{aligned}
    \frac{1}{N_\Delta}&\sum_{\mathbf{q}_i \in \text{bin } i}\delta^K(\mathbf{q}_{123}) B(\bb{q}_1,\bb{q}_2,\bb{q}_3) \\ &\approx \frac{1}{V_\Delta}\int_{k_{1,2,3}} \md^3\bb{q}_{1,2,3} \, \delta^D(\mathbf{q}_{123}) B(\bb{q}_1,\bb{q}_2,\bb{q}_3);
\end{aligned}
\end{equation}
where 
\begin{equation}
    V_\Delta = \int_{k_{1,2,3}} \md^3\bb{q}_{1,2,3} \, \delta^D(\mathbf{q}_{123}) 
\end{equation}
is the volume of the integration region.
This integration extends both over the angles that specify the orientation of the triangle as well as the `radial bin' that specifies the length of wavevectors. We will refer to this as the `integral approximation'\footnote{Crucially, we do not replace the volume $V_\Delta$ by $8\pi^2 k_1k_2k_3 (\Delta k)^3$, which is not accurate for very squeezed or folded configurations. Instead, the integral is calculated numerically - see also \cite{biagetti_squeezed} for exact analytic expressions.}. This approximation is valid in the limit of a sufficiently dense wavenumber grid (i.e. sufficiently high resolution). An additional common approximation involves replacing the radial bin integration in \eqref{eq:intapprox} with an evaluation of the (angular-averaged) integrand at effective wavenumbers:
\begin{equation}
\begin{aligned}
    \frac{1}{V_\Delta}&\int_{k_{1,2,3}} \md^3\bb{q}_{1,2,3}\,\delta^D(\mathbf{q}_{123})B(\bb{q}_1,\bb{q}_2,\bb{q}_3) \\
    &\approx \frac{1}{4\pi}\int \md \mu_1 \md \xi \,B(k^\text{eff}_1,k^\text{eff}_2,k^\text{eff}_3,\mu_1,\xi).
\end{aligned}
\end{equation}
\begin{figure}[t]
    \centering
    \includegraphics[width=\linewidth]{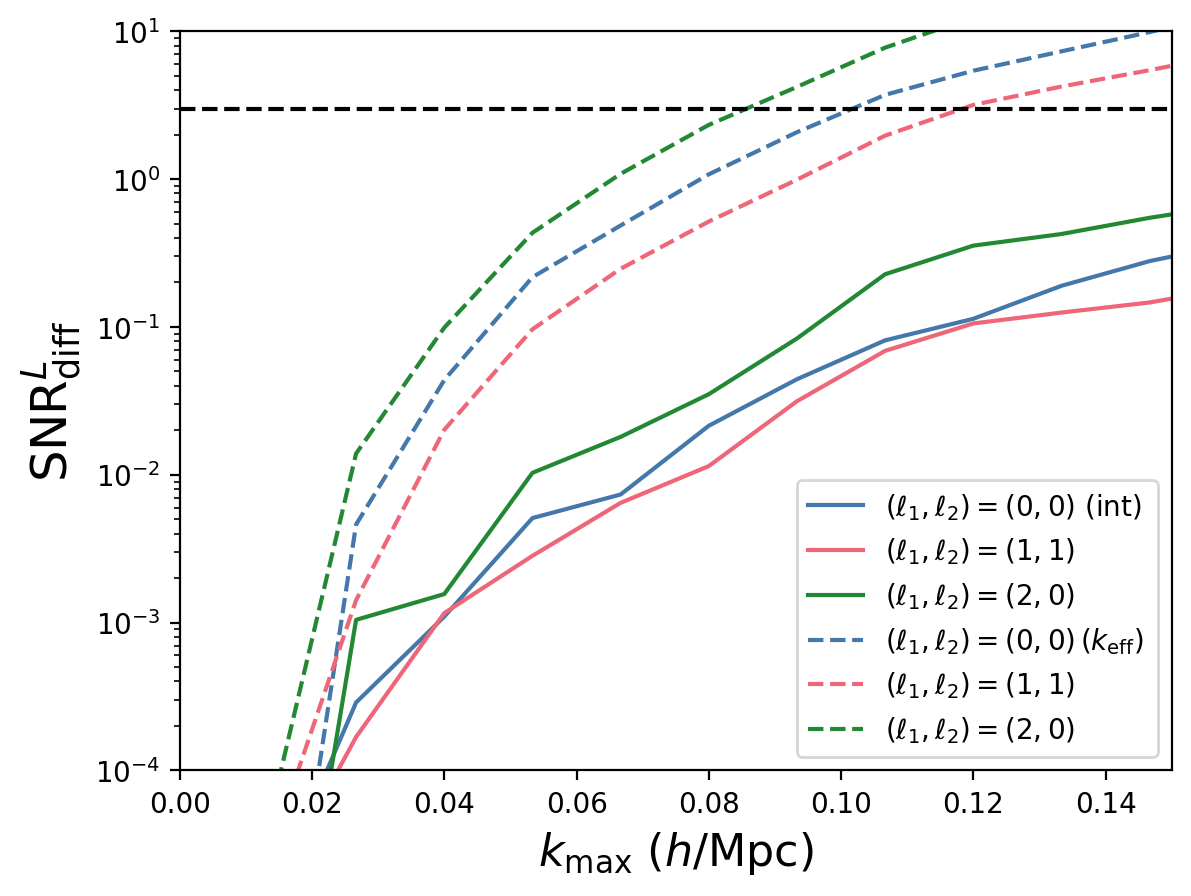} 
    \caption{SNR of the difference of the integral approximation (solid lines) and the effective wavenumber approximation (dashed lines) for three different multipoles of $B_{\delta \delta E}$. The black dashed horizontal line indicates an SNR of 3, which we take as a rough threshold for what is admissible.}
    \label{fig:keff_kint}
\end{figure}
We will refer to this as the `effective wavenumber approximation'. This approximation will break down if the radial bins are too wide, or if the bispectrum varies strongly in the radial direction. One must exercise caution when attempting to apply these approximations \cite{oddo_bisp}.
We tested their validity for our binning scheme by computing explicitly the sum over wavevectors and comparing to various choices for the wavenumbers $k_i^\text{eff}$ at which the bispectrum is evaluated. We also consider a comparison with numerical (Monte Carlo) integration over the wavevector bin, as done in e.g. \cite{damico_bisp, ivanov_bisp,philcox_bisp}. We then compute the SNR of the difference with respect to the explicit bin summation, using the diagonal sample variance as an error estimate, i.e. 
\begin{equation}
\begin{aligned}
    (\text{SNR}^L_\text{diff})^2 = \sum_{\text{bin } T}&(B_{\delta \delta E}^{L,\text{exact}}(T)-B_{\delta \delta E}^{L,\text{app.}}(T))C^{-1}_{LL}(T)\\
    &\times (B_{\delta \delta E}^{L,\text{exact}}(T)-B_{\delta \delta E}^{L,\text{app.}}(T)).
\end{aligned}
\end{equation}
Here the `app.' superscript refers to either the integral approximation or the effective wavenumber approximation and we do not consider cross-covariances between different multipoles. The result is shown in Figure \ref{fig:keff_kint}. We would like for this SNR to remain small, i.e. below unity.
We can clearly see that using the effective wavenumber approximation (dashed lines) is uncontrolled already at $k=0.10\,h/\text{Mpc}$. In contrast, the integral approximation (solid lines) fares much better, with the SNR of the difference staying below unity for all relevant scales. We decide to perform exact binning of the theoretical predictions, as also done in \cite{bakx_eft}. Here the situation may be more severe than for the scalar galaxy bispectrum, for which the multipoles exhibit weaker angular dependence, thus rendering the summand smoother and less susceptible to discreteness effects\footnote{It is also not clear whether the ordinary Legendre multipoles from Paper I would be less susceptible to discreteness effects; note that the summand in that case has a discontinuity along the $z$-axis where the phase factor $\exp 2\mi \phi_\bb{k}$ is undefined.}.

\subsection{Performance Metrics}
We assume that the likelihood $\mathcal{L}$ of the data is approximately Gaussian \cite{tucci}, i.e. the data vector follows a Gaussian distribution around a mean determined by the true theoretical model parameters. Then, suppressing the sum over triangles from now on, the log-likelihood is defined via the chi-squared statistic
\begin{equation}\label{eq:logl}
    \chi^2 = -2 \log \mathcal{L} =  \sum_{L,L'}^{k_\text{max}}(\hat{\bar{B}}_L-B_L^\text{th})C_{LL'}^{-1}(\hat{\bar{B}}_{L'}-B_{L'}^\text{th})
\end{equation}
where $L,L'$ are multipole indices, $C_{LL'}^{-1}$ is the precision matrix for a single realization and $\hat{\bar{B}}_L$ is the sample mean over all realizations as in Eq. \eqref{eq:covest}. We suppress the tracers $XYZ$ and the triangle bins for clarity. Thus, we carry out our fiducial analysis at $V = 1\, (\text{Gpc}/h)^3$ as in \cite{bakx_eft}. By using the mean of the data vectors over all realizations, the statistical noise becomes suppressed with respect to the covariance used. This allows us to more cleanly identify modeling systematics at a given scale cut. We will explore several different sets of multipoles to be included in the likelihood in Section \ref{sec:results}.

We recall that the power spectrum analysis of \cite{bakx_eft} inferred 
\begin{equation}\label{eq:b1fid}
    b_1^\text{g,fid} = -0.0776, \quad \sigma_1^\text{fid} = 0.002,
\end{equation}
from fitting the linear alignment model on large scales and
\begin{equation}\label{eq:b2fid}
\begin{aligned}
    b_{2,2}^\text{g,fid} &= 0.129, \quad \sigma_{2,2}^\text{fid} = 0.008; \\
    b_{2,3}^\text{g,fid} &= -0.054, \quad \sigma_{2,3}^\text{fid} = 0.017;
\end{aligned}
\end{equation}
from the EFT fit in Table 2 of \cite{bakx_eft}. Establishing consistency of the bispectrum with this measurement will be a key focus of our work. Thus, as another performance metric we will consider the (Gaussian) figure of bias (FoB), i.e. 
\begin{equation}\label{eq:fob}
    \text{FoB}(k_\text{max}) = \left[(\vec{b}^\text{g}(k_\text{max})-\vec{b}^\text{g,fid})S_\text{tot}^{-1}(\vec{b}^\text{g}(k_\text{max})-\vec{b}^\text{g,fid})\right]^{\frac{1}{2}}
\end{equation}
where 
\begin{equation}\label{eq:fobcov}
    S_\text{tot} = S(k_\text{max})+\text{diag}\big((\sigma_1^\text{fid})^2,(\sigma_{2,2}^\text{fid})^2,(\sigma_{2,3}^\text{fid})^2\big)
\end{equation}
is the posterior covariance of the difference of the bias vectors. Here $\vec{b}$ is the vector of biases $(b_1^\text{g},b_{2,2}^\text{g},b_{2,3}^\text{g})$ and $S(k_\text{max})$ is the bispectrum posterior covariance of these three variables. The second term in Eq. \eqref{eq:fobcov} amounts to adding the power spectrum posterior covariances of the diagonal\footnote{For simplicity we do not consider posterior correlations between biases inferred in \cite{bakx_eft}, but only add the diagonal covariance.}. The FoB quantifies how well the fiducial bias parameters are recovered by the bispectrum. The 68 (95) percent confidence level for Eq. \eqref{eq:fob} equals 1.88 (2.83) - note that this value depends on the number of variables. We do not include $b_{2,1}^\text{g}$ in this metric since its posterior was bimodal in the analysis of \cite{bakx_eft}. Nevertheless, we did obtain relevant constraints on it; one can see in Figure 4 of \cite{bakx_eft} that e.g. $|b_{2,1}^\text{g}| < 0.04$ with high confidence. As we will see, this requirement is also satisfied by the bispectrum posteriors.

We also assess the goodness-of-fit of the model; following \cite{egge_pow} we consider the chi-squared statistic averaged over all 20 realizations, i.e. 
\begin{equation}\label{eq:chisqbar}
    \bar{\chi}^2 = \frac{1}{N_r} \sum_{L,L',r}^{k_\text{max}}(\hat{B}_{r,L}-B_{L}^\text{th})C_{LL'}^{-1}(\hat{B}_{r,L'}-B_{L'}^\text{th}).
\end{equation}
Assuming that (i) the best-fit model $B_{L,*}^\text{th}$ found from minimizing $\bar{\chi}^2$ is the true mean of the distribution of $\hat{B}$ and (ii) the precision matrix $C_{LL'}^{-1}$ equals the true precision matrix, the variable $\bar{\chi}^2_* = \bar{\chi}^2(B_{L,*}^\text{th})$ follows a $\chi^2$ distribution, which is approximately normal\footnote{From now on, we will omit the asterisk on chi-squared statistics to avoid clutter. When referring to the distribution of e.g. Eq. \eqref{eq:chisqbar}, it is understood that it is evaluated at the best-fit value. For a more complete treatment of how this can impact e.g. the number of degrees of freedom, see \cite{raveri}.}. When the covariance (and hence the precision matrix) is known exactly and one ignores the number of free parameters, which is much less than $N_L N_T$ (where $N_T$ is the number of triangle bins), one recovers the standard result 
\begin{equation}
    \bar{\chi}^2 \sim \mathcal{N}(\mu = N_L N_T, \sigma^2 = 2 N_L N_T)
\end{equation}
which after dividing by the number of degrees of freedom $N_L N_T$ reads 
\begin{equation}\label{eq:chisqdist}
    \chi_\text{red}^2 = \frac{\bar{\chi}^2}{N_LN_T} \sim \mathcal{N}(\mu_\text{red} = 1, \sigma_\text{red}^2 = 2/(N_LN_T)).
\end{equation}
Here $\mathcal{N}(\mu, \sigma^2)$ denotes a normal distribution. Thus, we can assess the goodness of fit by checking whether the left-hand side is sufficiently close to unity for the best-fit model. If we assume the Gaussian covariance to be the true one, we can apply Eq. \eqref{eq:chisqdist} without further issues. However, here an additional subtlety arises when the covariance is \textit{estimated from the same data}, as with Eq. \eqref{eq:covest}. In that case, the mean $\hat{\mu}_\text{red}$ and variance $\hat{\sigma}_\text{red}^2$ of $\hat{\bar{\chi}}^2/(N_T N_L)$ (where the hat indicates that Eq. \eqref{eq:covest} is used for the covariance matrix) are \textit{not} equal to the values from Eq. \eqref{eq:chisqdist}, \textit{even after applying the Hartlap correction from Eq. \eqref{eq:hartlapfac}}. Moreover, \textit{$\hat{\sigma}_\mathrm{red}^2$ does not approach $\sigma_\mathrm{red}^2$ even when $N_r \to \infty$}. We elaborate on this issue in App. \ref{sec:cov}, where we also derive the correct expressions for the mean and standard deviation of $\hat{\bar{\chi}}^2$ in this specific case. The chi-squared test can then still be used to assess the goodness of fit\footnote{Strictly speaking, there are additional complications when using a noisy estimate for the precision matrix. In particular, the noise in the covariance matrix needs to be propagated into the likelihood inference as well \cite{sellentin}. We ignore these issues here.}.  

Lastly, to facilitate a direct comparison to \cite{bakx_eft}, we consider a metric comprised of a weighted sum of $\chi_\text{red}^2$ and the FoB\footnote{We take the absolute value of the second term in this definition in order to penalize the scenario where a large FoB is compensated by overfitting (where the second term without absolute values would be negative).}
\begin{equation}\label{eq:delta}
    \Delta(k_\text{max}) = \frac{\text{FoB}}{1.88} + \frac{|\chi_\text{red}^2-\mu_\text{red}|}{2\sigma_\text{red}}.
\end{equation} 
We will compute all the corresponding quantities defined above in both the case where the Gaussian covariance is used, and the case where Eq. \eqref{eq:covest} is used, taking into account the above discussion. In order to be roughly consistent with \cite{bakx_eft}, the FoB is normalized by its 68 percent confidence level. 
In \cite{bakx_eft}, a threshold of $\Delta_\text{crit} = 1.5$ was used to define the breakdown of the model (see also \cite{egge_pow,egge_joint}). In this work, given the uncertainties we face when computing the $\chi^2$ statistic\footnote{This is due to the uncertainty in the bispectrum covariance; the large number of degrees of freedom renders the p-value for the $\chi_\text{red}^2$ statistic sensitive to even $\sim 5\%$ differences in the covariance.} we opt to be somewhat more lenient and use $\Delta_\text{crit} = 2$. 
\begin{figure}[t]
    \centering
    \includegraphics[width=\linewidth]{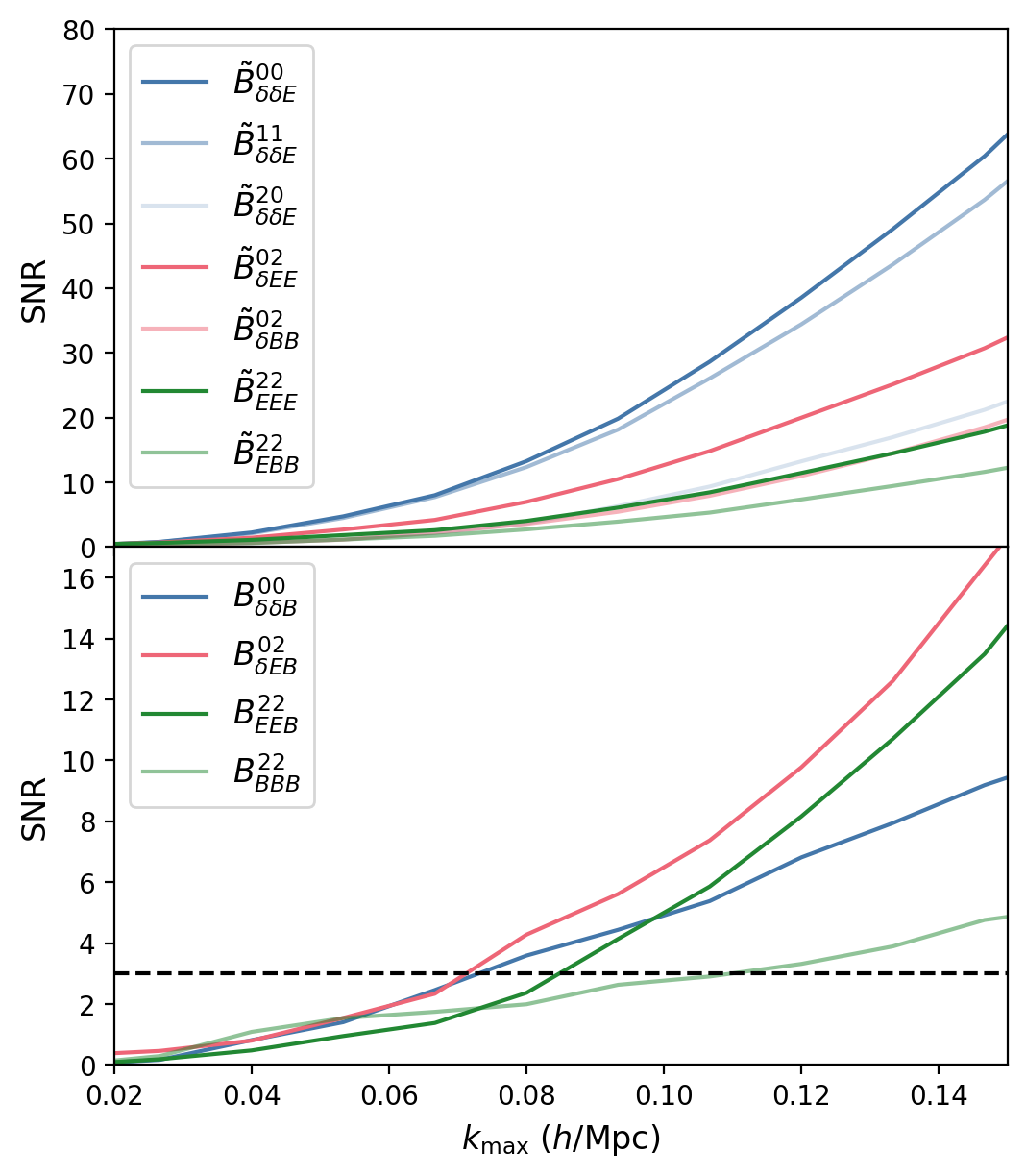}
    \caption{\textit{Top Panel}: SNR for the lowest order multipoles of the parity-even bispectra from Eq. \eqref{eq:combs} (with the exception of the matter bispectrum), as well as the two higher order multipoles of $B_{\delta \delta E}$ that were also considered in Paper I. \textit{Bottom Panel}: same as top panel but for the parity-odd 
    bispectra. The black dashed line indicates an SNR of 3.}
    \label{fig:snreven}
\end{figure}
As emphasized before, in order to maintain consistency with \cite{bakx_eft} we perform our baseline analysis for the volume of a single realization, i.e. $V=1\,(\text{Gpc}/h)^3$; that is, we fit the mean data vector and use a covariance corresponding to $V=1\,(\text{Gpc}/h)^3$ (see Eq. \eqref{eq:logl}). When performing an MCMC analysis, uninformative uniform priors of $[-0.5,0.5]$ on the linear shape bias $b_1^\text{g}$, $[-3,3]$ on the second order bias parameters and $[-30,30]$ on the stochastic parameter $\alpha_\text{gg}^{1\delta}$. In the Poisson limit we would expect this parameter to be equal to 
\begin{equation}\label{eq:pois}
    \sigma_\gamma^2/2\bar{n} \approx 4.6\,(\text{Mpc}/h)^3,
\end{equation}
where $\sigma_{\gamma}^2 \approx 0.2$ is the total RMS ellipticity and $\bar{n} \approx 3 \times 10^{-3} (h/\text{Mpc})^3$ is the number density of halos. We used the affine invariant Metropolis-Hastings sampler implemented in \texttt{emcee} \cite{emcee} with 32 walkers and $50,000$ steps, discarding the first $100$ burn-in steps. The posteriors are post-processed using \texttt{GetDist} \cite{getdist}. 
\section{Results}\label{sec:results}
Before discussing the fit to the EFT of IA, we first present the detection significance of our bispectrum multipole measurements for a single realization volume, using scales up to $k_\text{max} = 0.15\,h/\text{Mpc}$. That is, we calculate 
\begin{equation}
    (\text{SNR}^L_{XYZ})^2 = \sum_{\text{bin } T}\hat{\bar{B}}_{XYZ}^{L}(T)C^{-1}_{LL}(T)\hat{\bar{B}}_{XYZ}^{L}(T),
\end{equation}
where $\hat{\bar{B}}_{XYZ}^{L}(T)$ is the mean of the measurements over all realizations.
\begin{figure}[t]
    \centering
    \includegraphics[width=\linewidth]{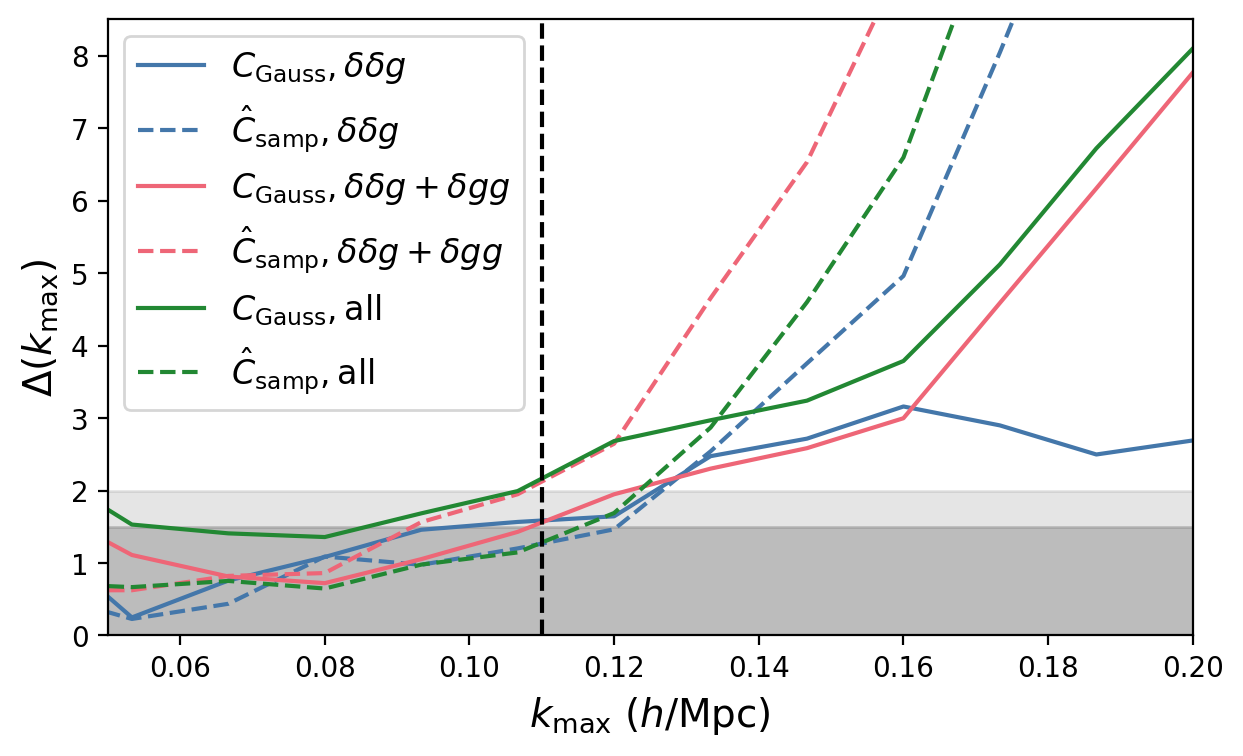}
    \caption{Results for the breakdown criterion from Eq. \eqref{eq:delta} for all different cases considered in Section \ref{sec:results}. Solid lines indicate the use of the theoretical Gaussian covariance, while dashed lines indicate the use of the `block diagonal sample variance' from Eq. \eqref{eq:covest}. Blue, pink and green refer to at most one, two or three shape fields, as discussed in Sections \ref{subsec:onefield}, \ref{subsec:twofield} and \ref{subsec:threefield} respectively. The shades indicate $\Delta = 1.5$ and $\Delta = 2$ respectively and the vertical dotted line indicates $k_\text{max} = 0.11\,h/\text{Mpc}$, the maximum wavenumber at which we deem the tree-level EFT of IA model to be a good fit to the data.}
    \label{fig:deltaall}
\end{figure}
We used the diagonal sample variance to compute the SNR, but found no appreciable difference when using the Gaussian covariance instead. The result is shown in Figure \ref{fig:snreven}, where we clearly see in the top panel that all parity-even multipoles are detected at $k_\text{max} = 0.11\,h/\text{Mpc}$ (which will turn out to be the wavenumber at which our model breaks down), with the monopole $\tilde{B}^{00}_{\delta \delta E}$ having the highest SNR. All parity-odd multipoles are also detected at this wavenumber, except for $B_{BBB}^{22}$, which we indeed expect to be zero at tree-level in the EFT of IA (see the discussion around Eq. \eqref{eq:bmodezero}). This also serves as an important consistency check for our parity-odd estimators; the inclusion of the cross-product in Eq. \eqref{eq:asproject} allows us to extract the 3D bispectrum signal after averaging over all triangle orientations. It is important to note that these SNR values are obtained without performing any kind of `noise' subtraction from the measured bispectrum, that is, the stochastic contributions proportional to $\alpha_\text{gg}^{1\delta}$ in $B^{\delta \text{gg}}$ and $B^\text{ggg}$ are not removed. These contributions constitute a large part of the measured bispectra, and we will come back to this issue in Sections \ref{subsec:twofield} and \ref{subsec:threefield}. Importantly (and in line with general expectation), the SNR does not directly translate into the quality of parameter constraints.
\begin{figure}[t]
    \centering
    \includegraphics[width=\linewidth]{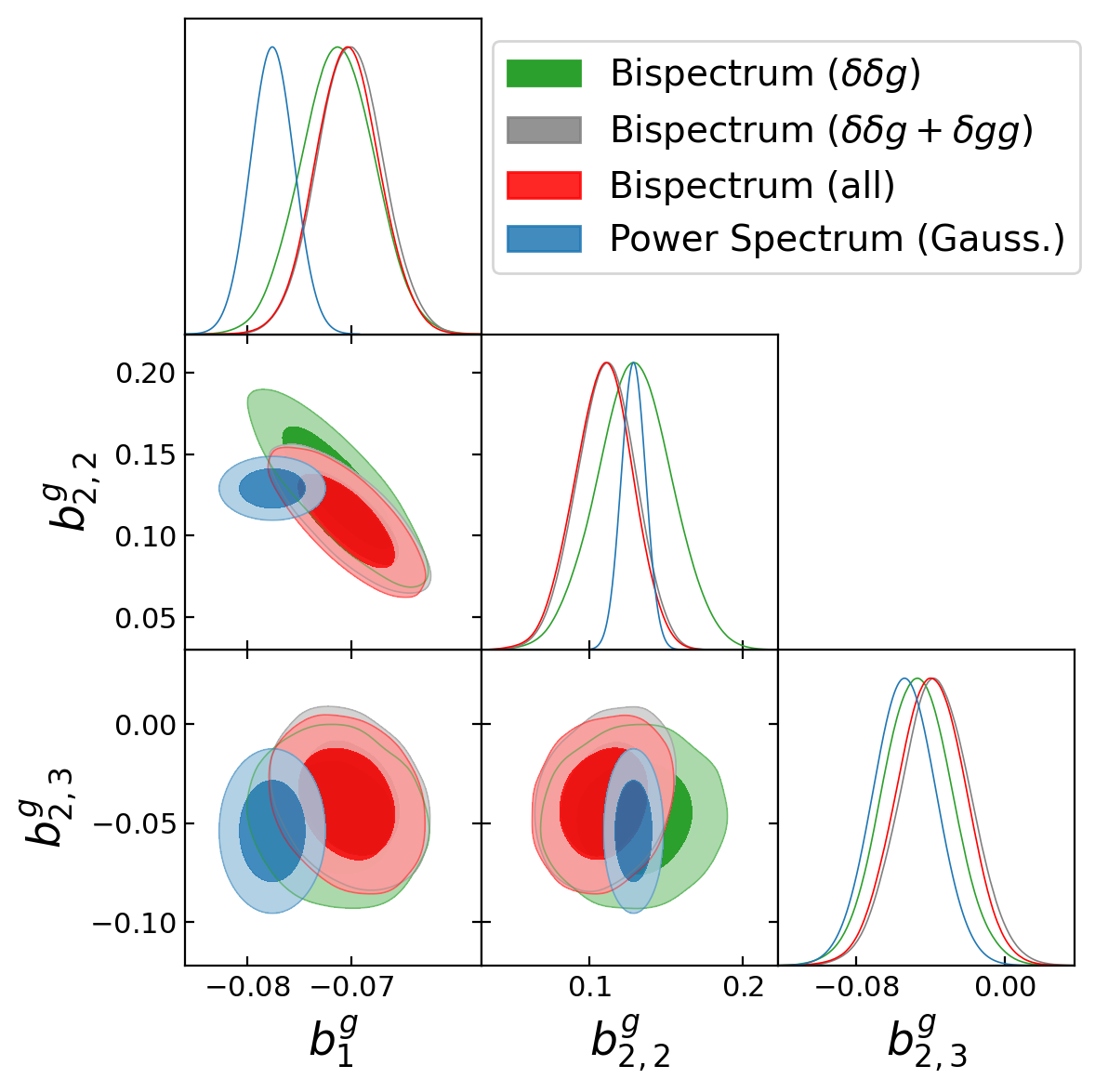}
    \caption{Comparison of the posteriors from bispectra with at most one (green), two (grey) or three (red) shape fields at $k_\text{max} = 0.11\,h/\text{Mpc}$. Also show is the (Gaussian) power spectrum posterior from \cite{bakx_eft}, cf. Eq. \eqref{eq:b1fid} and \eqref{eq:b2fid} (where a prior on the sign of $b_{2,2}^\text{g}$ and $b_{2,3}^\text{g}$ had to be applied). Observe that the grey and red posteriors are almost identical. Here we used the Hartlap-corrected diagonal sample variance from Eq. \eqref{eq:covest}}.
    \label{fig:corner_all}
\end{figure}
In our most extensive analysis, we include seven parity-even multipoles in the likelihood, 

\begin{equation}\label{eq:liks}
\begin{aligned}
    \delta \delta \text{g}: &\quad \{\tilde{B}_{\delta \delta E}^{00},\tilde{B}_{\delta \delta E}^{11}, \tilde{B}_{\delta \delta E}^{20}\}, \\ 
    \delta \text{g} \text{g}: &\quad \{\tilde{B}_{\delta E E}^{02},\tilde{B}_{\delta B B}^{02}\}, \\
    \text{g} \text{g} \text{g}: &\quad \{\tilde{B}_{EE E}^{22},\tilde{B}_{EBB}^{22}\},
\end{aligned}
\end{equation}
\begin{figure}[t]
    \centering
    \includegraphics[width=\linewidth]{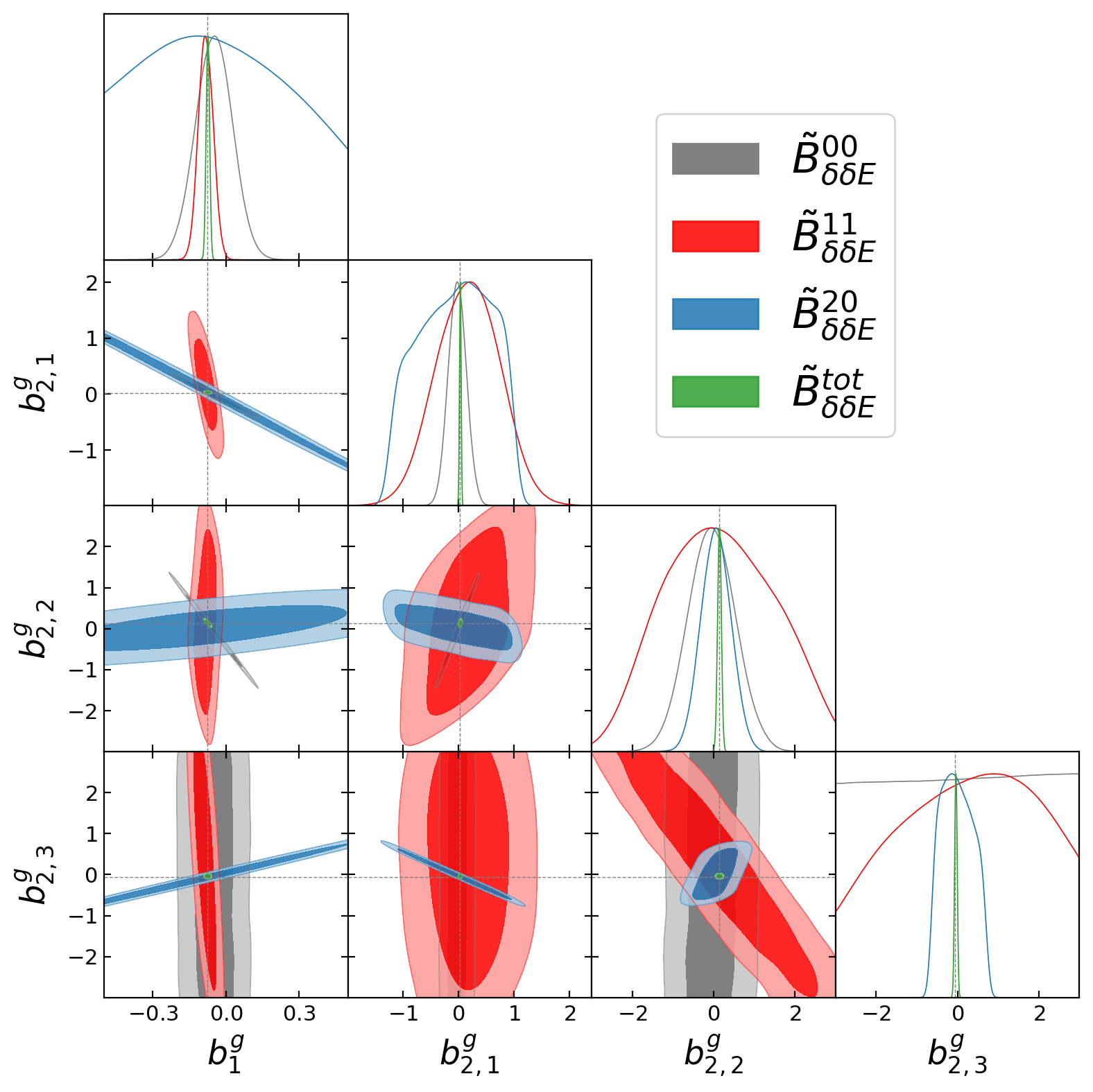}
    \caption{Posteriors for the first and second order IA bias parameters at $k_\text{max} = 0.11\,h/\text{Mpc}$ when analyzing individual multipoles of $B_{\delta \delta E}$ and when combining them. It is apparent that each multipole individually suffers from severe degeneracies, but combining them (blue contours) yields very good constraints on their amplitudes. Here we used the Hartlap-corrected diagonal sample variance from Eq. \eqref{eq:covest}.}
    \label{fig:degenplot}
\end{figure}
\begin{figure}[t]
    \centering
    \includegraphics[width=0.95\linewidth]{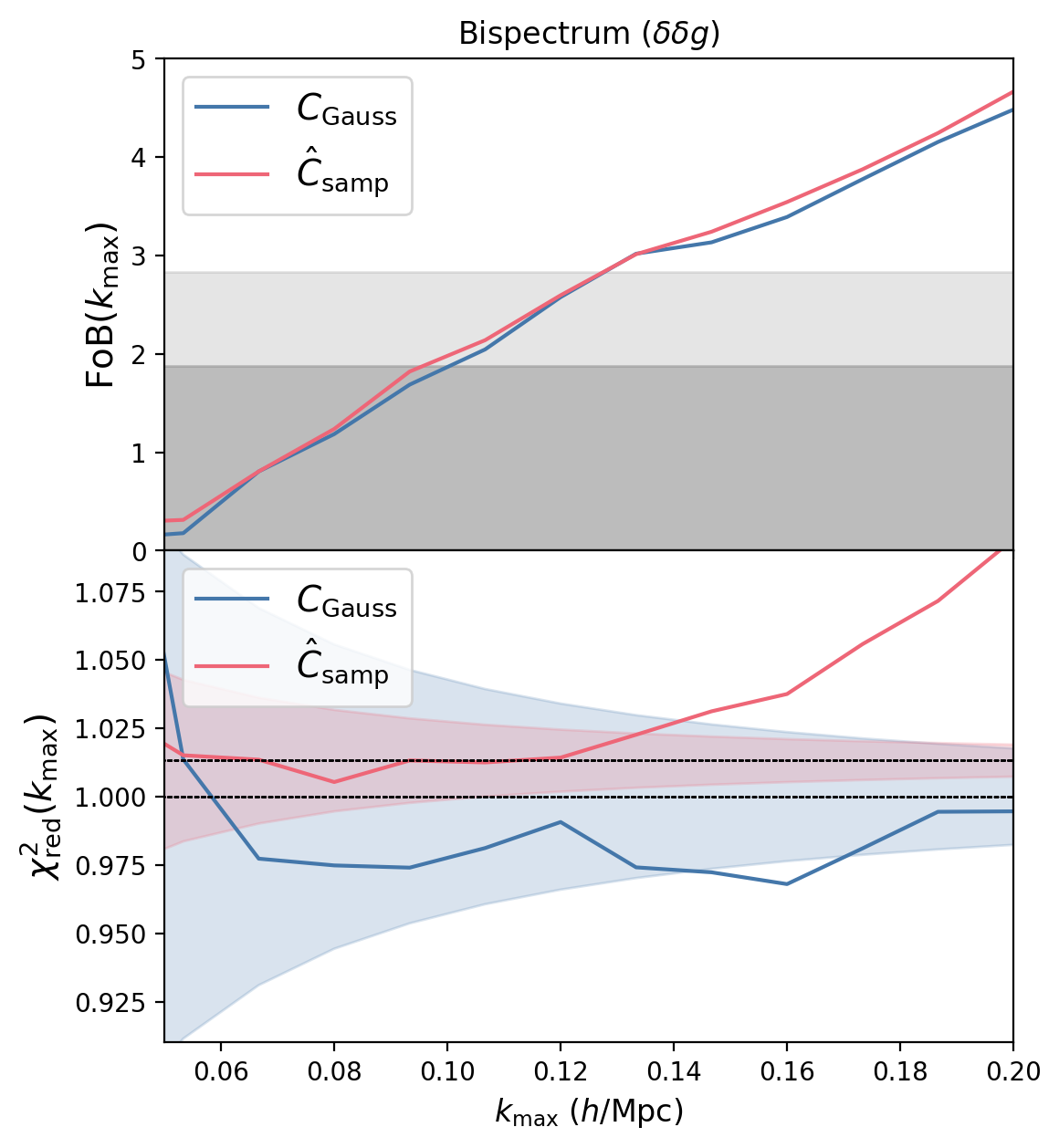}
    \caption{FoB and $\chi_\text{red}^2$ for the Gaussian (blue) and diagonal sample covariance (pink) for the likelihood including bispectra with one shape field. The shaded regions indicate confidence intervals; in the top panel we plot the $1\sigma$ ($2\sigma$) confidence intervals in dark (light) grey for the FoB. In the bottom panel, the blue shades indicate the $2\sigma$ confidence intervals $[\mu_\text{red} - 2\sigma_\text{red}, \mu_\text{red} + 2\sigma_\text{red}]$ and the pink shades $[\hat{\mu}_\text{red} - 2\hat{\sigma}_\text{red}, \hat{\mu}_\text{red} + 2\hat{\sigma}_\text{red}]$ (see the discussion around Eq. \eqref{eq:chisqdist}) and the dashed black lines indicate the values of $\mu_\text{red} = 1$ and $\hat{\mu}_\text{red}$ as in Eq. \eqref{eq:chisqresult}.}
    \label{fig:fitplot_sst}
\end{figure}
\begin{figure*}[t]
    \centering
    \includegraphics[width=\linewidth]{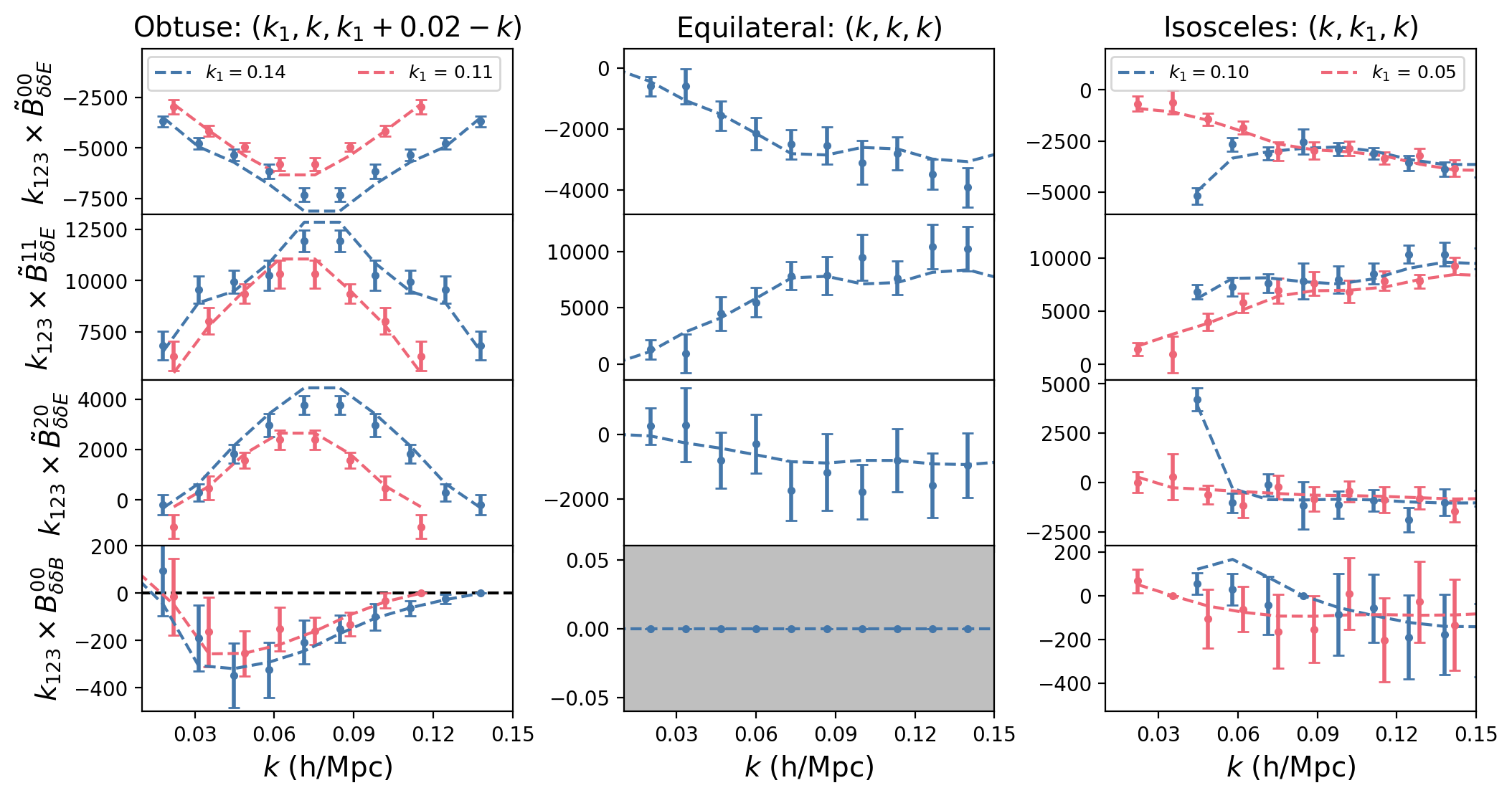}
    \caption{Measured bispectrum signal for the four multipoles with one tensor field considered in this work. We plot the product $k_1k_2k_3\times B$ in units of $(\text{Mpc}/h)^3$. Each row contains a specific multipole, while each column contains a specific type of triangle configuration. In the left column, we plot `obtuse' configurations where the sum of the two shortest sides exceeds the longest by roughly $0.02\, h/\text{Mpc}$. Note that these plots are therefore symmetric about the `middle bin' for the even multipoles (first three rows); that is, the same data is plotted twice, but this is not true for $B_{\delta \delta B}^{00}$. In the leftmost and rightmost columns, we choose two specific values for the remaining `fixed' side $k_1$ (pink and blue). The errors have been rescaled to match a volume of $V=7.5\,(\text{Gpc}/h)^3$. The dashed lines show the theory prediction using the values from Eq. \eqref{eq:plotparams}. The bispectra have been multiplied by $k_{123} = k_1 k_2 k_3$ for visualization purposes.}
    \label{fig:sigplot_sst}
\end{figure*}
\noindent where the respective lines indicate bispectra with one, two or three tensor fields (cf. Eq. \eqref{eq:combs}). Note that for the second line, the $(\ell_1,\ell_2) = (0,2)$ multipole is in fact the lowest order non-vanishing one, and a similar remark applies to the third line. We will discuss the impact of adding each line to the likelihood extensively in the three subsections below, i.e. Section \ref{subsec:onefield} deals with bispectra with one shape field, while Section \ref{subsec:twofield} deals with the five bispectra with at most two shape fields and Section \ref{subsec:threefield} discusses the likelihood with all seven bispectra above. All fits are shown both for the use of the (Hartlap-corrected) diagonal sample variance from Eq. \eqref{eq:covest} as well as the theoretical Gaussian covariance. Our main results are summarized in Figures \ref{fig:deltaall} and \ref{fig:corner_all}. In Figure \ref{fig:deltaall}, we see that the bispectrum is nominally well described by the EFT of IA up to $k_\text{max} = 0.11\,h/\text{Mpc}$ (vertical dashed line), with some variations depending on the specific setup. For $k_\text{max} = 0.11\,h/\text{Mpc}$, the posteriors on the three parameters included in the FoB metric from Eq. \eqref{eq:fob} are plotted in Figure \ref{fig:corner_all}, along with the power spectrum result from \cite{bakx_eft}, indicating good consistency between two- and three-point statistics. The bispectrum posteriors are still broader than those from the power spectrum, but given that the power spectrum analysis used a much larger value of $k_\text{max} = 0.28 \, h/\text{Mpc}$ this is not surprising. 
\subsection{One shape field}\label{subsec:onefield}
We begin by considering multipoles of $B_{\delta \delta E}$ in the top line of Eq. \eqref{eq:liks}. The lowest order multipole $\tilde{B}_{\delta\delta E}^{00}$ does not depend on $b_{2,3}^\text{g}$. Therefore, in order to constrain all bias parameters we need to include higher order multipoles as well. The degeneracy is broken extremely efficiently by the two quadrupoles $\tilde{B}_{\delta\delta E}^{20},\tilde{B}_{\delta\delta E}^{11}$, and Figure \ref{fig:degenplot} illustrates this. 

When analyzing them jointly, we obtain the green posterior shown in Figure \ref{fig:corner_all} at a scale cut of $k_\text{max} \approx 0.11 \, h/\text{Mpc}$, which for reference also overplots the (Gaussian) power spectrum posteriors from Eqs. \eqref{eq:b1fid} and \eqref{eq:b2fid}. We find good consistency between the inferred parameter values of $b_1^\text{g}, b_{2,2}^\text{g}$ and $b_{2,3}^\text{g}$ and we also check that the $b_{2,1}^\text{g}$ posterior from the bispectrum is compatible with the one for the power spectrum. Specifically, we obtain
\begin{equation}\label{eq:b21res}
    b_{2,1}^\text{g} = 0.031 \pm 0.009.
\end{equation}
The bispectrum seems to prefer somewhat lower values of $b_1^\text{g}$ than the power spectrum analysis, although this difference is small for this range of scales. Moreover, in the absence of any knowledge on the cross-covariance between the power spectrum and bispectrum likelihoods, it is also possible that using Eq. \eqref{eq:fob} in its current form is in fact overestimating the tension; this would be the case if the power spectrum and bispectrum are anti-correlated. We leave a more thorough investigation of this aspect to future work.

The performance metrics discussed in Section \ref{sec:sims} are shown in Figure \ref{fig:fitplot_sst}, both for the Gaussian covariance $C_\text{gauss}$ and $\hat{C}_\text{samp}$ from Eq. \eqref{eq:covest}. We observe good consistency in the FoB between the two choices. However, in the case of $\chi_\text{red}^2$ we see that the Gaussian covariance yields systematically lower values. This does not impact the reach of the model much according to the $\Delta(k_\text{max})$ criterion, as seen by examining the two blue lines in Figure \ref{fig:deltaall}. Moreover, we also checked that for $k_\text{max} \gtrsim 0.14\, h/\text{Mpc} $ some of the bias parameters change values significantly, which is a sign of overfitting since the values of the bias parameters should be uniquely determined. 
Altogether, we conclude that the three lowest order multipoles of $B_{\delta \delta E}$ are well fit by the tree-level EFT of IA up to at least $k_\text{max} \sim 0.11  \, h/\text{Mpc}$ (and likely somewhat beyond) and in good agreement with the power spectrum. 
\begin{figure*}[t]
    \centering
    \includegraphics[width=\linewidth]{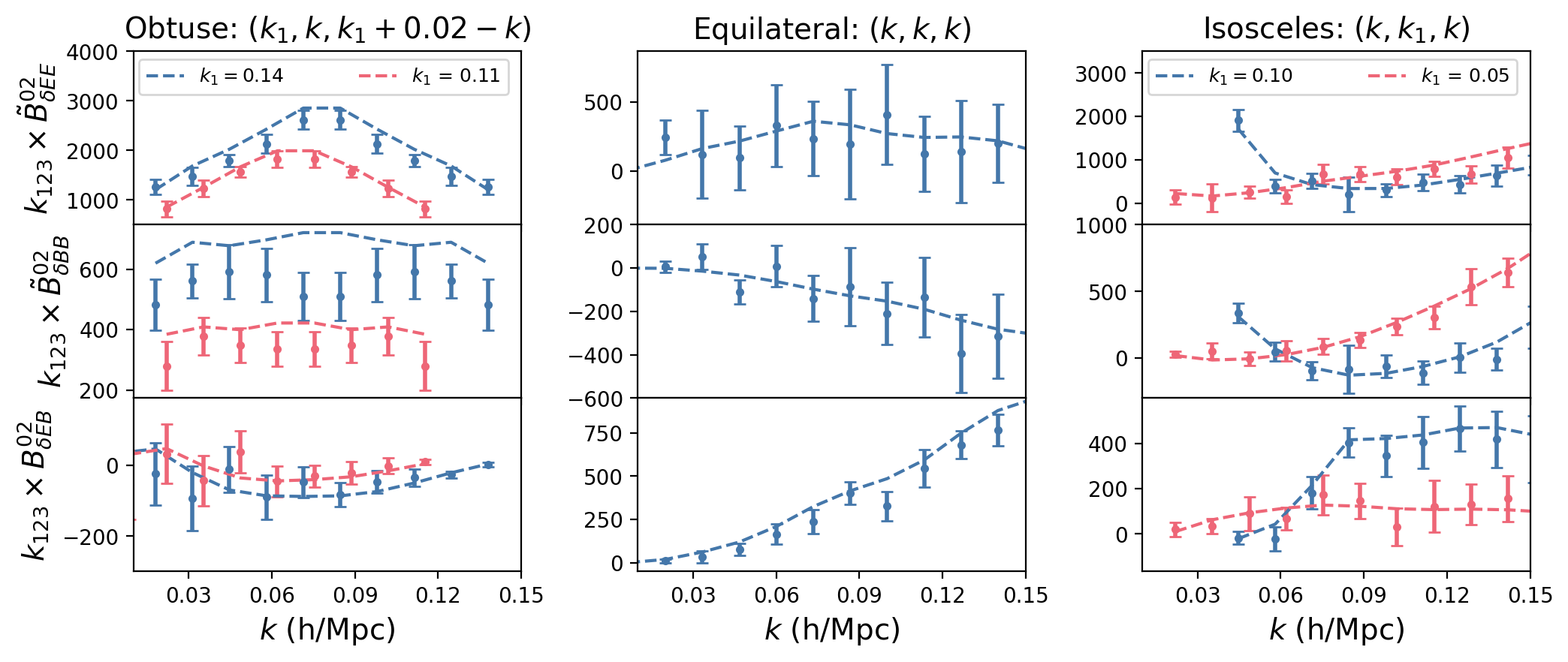}
    \caption{Same as Figure \ref{fig:sigplot_sst} but now for multipoles with two shape fields. The errors have been rescaled to match a volume of $V=7.5\,(\text{Gpc}/h)^3$. The dashed lines show the theory prediction using the values from Eq. \eqref{eq:plotparams}. The bispectra have been multiplied by $k_{123} = k_1 k_2 k_3$ for visualization purposes.}
    \label{fig:sigplot_stt}
\end{figure*}

Lastly, we also investigated the monopole of $B_{\delta \delta B}$ and observe that it is in good agreement with what is expected from $B_{\delta \delta E}$; note that $B_{\delta \delta B}^{00}$ only depends on $b_{2,1}^\text{g}$ as mentioned in Paper I. In all our plots of the measured signal, we show error bars corresponding to the total volume $V_\text{DESI} = 7.5\,(\text{Gpc}/h)^3$ of the DESI LRG sample\footnote{Note that this is $\sim 3\times$ larger than the volume considered in Paper I, where we assumed a smaller sky fraction due to possibly limited overlap with LSST.}. More concretely, Figure \ref{fig:sigplot_sst} shows the measured bispectrum signal\footnote{Note that in the equilateral (middle) panel, both the signal and the error bar are \textit{exactly} zero (that is, to machine precision) for odd bispectra, due to symmetry in the equilateral configuration. To see this, note that if bin 1 and bin 2 are the same and $X=Y$ in Eq. \eqref{eq:est}, swapping $\bb{q}_1$ and $\bb{q}_2$ in the sum leaves the summand invariant, but the cross product acquires a minus sign, so that the estimator equals minus itself, and hence equals zero. Thus, the covariance is also zero in this case.} for all four multipoles considered in this subsection, with the error bars rescaled to $V_\text{DESI} = 7.5\,(\text{Gpc}/h)^3$. We also overplot the theory curves using the parameter choices
\begin{equation}\label{eq:plotparams}
    \begin{aligned}
        b_{1}^\text{g} &= -0.070,\\
        b_{2,1}^\text{g} &= 0.020,\\
        b_{2,2}^\text{g} &= 0.11,\\
        b_{2,3}^\text{g} &= -0.040.
    \end{aligned}
\end{equation}
These are the best-fit model parameters from the joint fit of all multipoles (cf. Section \ref{subsec:threefield}). Even for this larger volume, the model is broadly in agreement with the data. However, one can also see that the equilateral $\tilde{B}_{\delta \delta E}$ bispectrum does not match beyond $k \sim 0.10 \,h/\text{Mpc}$ for $(\ell_1,\ell_2) = (0,0)$ and $(\ell_1,\ell_2) = (1,1)$ and `folded' configurations of the form $(k,k/2,k/2)$ are also poorly fit (this is also the region where $\tilde{B}_{\delta \delta E}$ has the highest SNR, so it is perhaps not surprising that the model fails there sooner). Observe that the parity-odd multipole $B_{\delta \delta B}^{00}$ is clearly nonzero for the obtuse configurations, but vanishes for equilateral triangles because of symmetries (see Figure 2 in Paper I).
\subsection{One or two shape fields}\label{subsec:twofield}
For bispectra with two tensor fields, i.e. $B_{\delta EE}, B_{\delta EB}$ and $B_{\delta BB}$, an additional parameter enters the tree-level model. This contribution was first derived in \cite{vlah_eft1} and labelled as $\alpha_\text{gg}^{1\delta}$ in Paper I. 
\begin{figure}[t]
    \centering
    \includegraphics[width=0.95\linewidth]{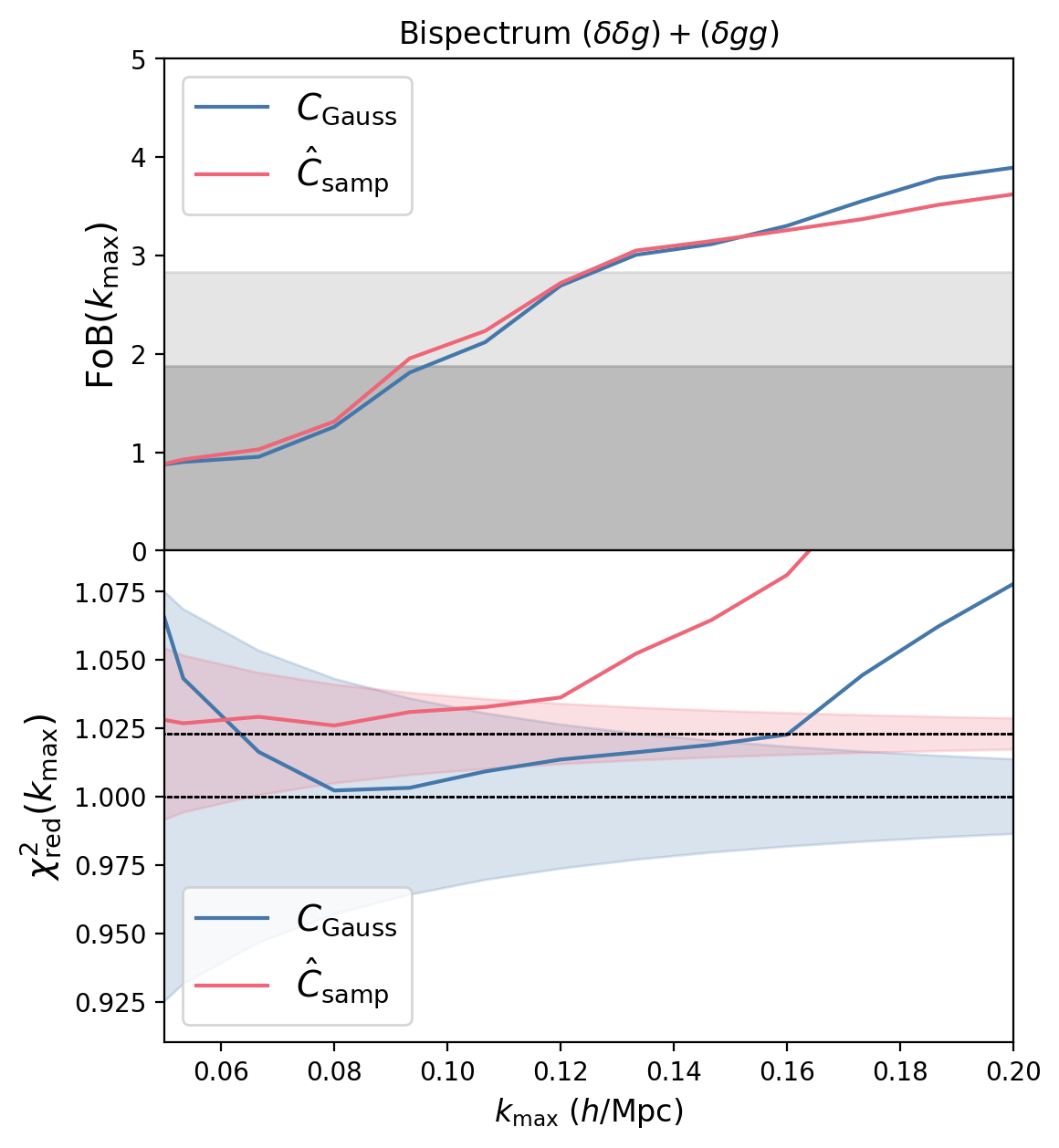}
    \caption{FoB and $\chi_\text{red}^2$ for the Gaussian (blue) and diagonal sample covariance (pink) for the likelihood including bispectra with one or two shape fields (cf. Figure \ref{fig:fitplot_sst}).}
    \label{fig:fitplot_stt}
\end{figure}
Interestingly, since the B-mode starts at second order, $B_{\delta BB}$ depends only on this parameter and not on any deterministic alignment bias (see also Table \ref{tab:params}). We measure the first nonvanishing multipoles of these bispectra, that is $(\ell_1,\ell_2) = (0,2)$. As explained before, we proceed by adding the parity-even bispectra $\tilde{B}^{02}_{\delta EE}$ and $\tilde{B}^{02}_{\delta BB}$ to the baseline likelihood (which thus consists of five multipoles now) and only check consistency with the parity-odd bispectrum $B^{02}_{\delta EB}$ at a later stage. The grey posteriors for $b_1^\text{g}, b_{2,2}^\text{g}$ and $b_{2,3}^\text{g}$ in Figure \ref{fig:corner_all} for $k_\text{max} \approx 0.11 \, h/\text{Mpc}$ again show good consistency, while the bispectrum posteriors shrink noticeably compared to the case with only one tensor field. Notably, both $B_{\delta EE}$ and $B_{\delta BB}$ are well described by using the same stochastic amplitude $\alpha_\text{gg}^{1\delta}$. Specifically, we obtain 
\begin{figure*}[t]
    \centering
    \includegraphics[width=\linewidth]{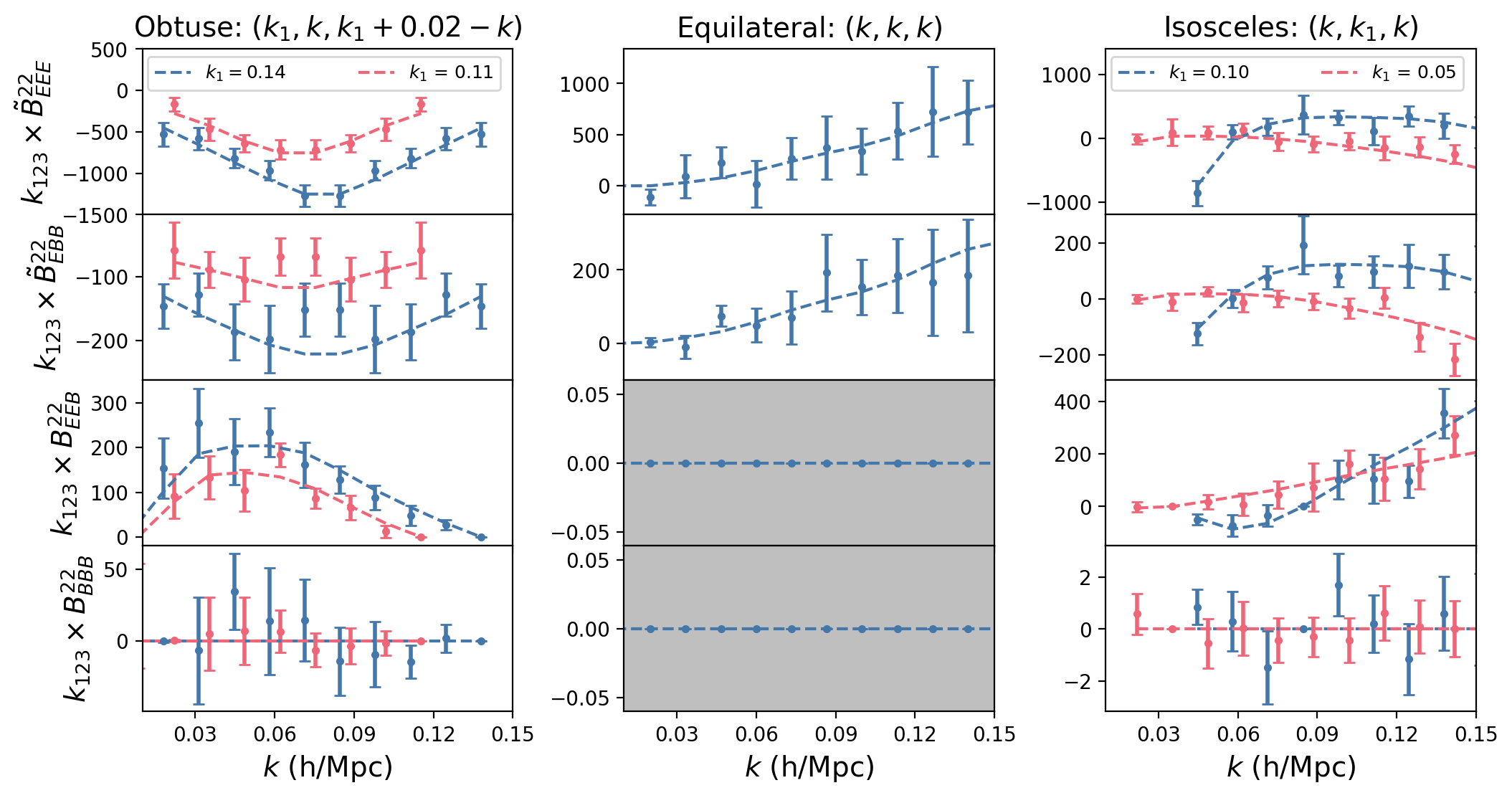}
    \caption{Same as Figure \ref{fig:sigplot_sst} but now for multipoles with three shape fields. Observe again that the odd bispectra vanish in the equilateral configuration (Figure 6 of Paper I) because of symmetries (and the fact that two of the fields are the same). The errors have been rescaled to match a volume of $V=7.5\,(\text{Gpc}/h)^3$. The dashed lines show the theory prediction using the values from Eq. \eqref{eq:plotparams}. The bispectra have been multiplied by $k_{123} = k_1 k_2 k_3$ for visualization purposes.}
    \label{fig:sigplot_ttt}
\end{figure*}
\begin{figure}[t]
    \centering
    \includegraphics[width=\linewidth]{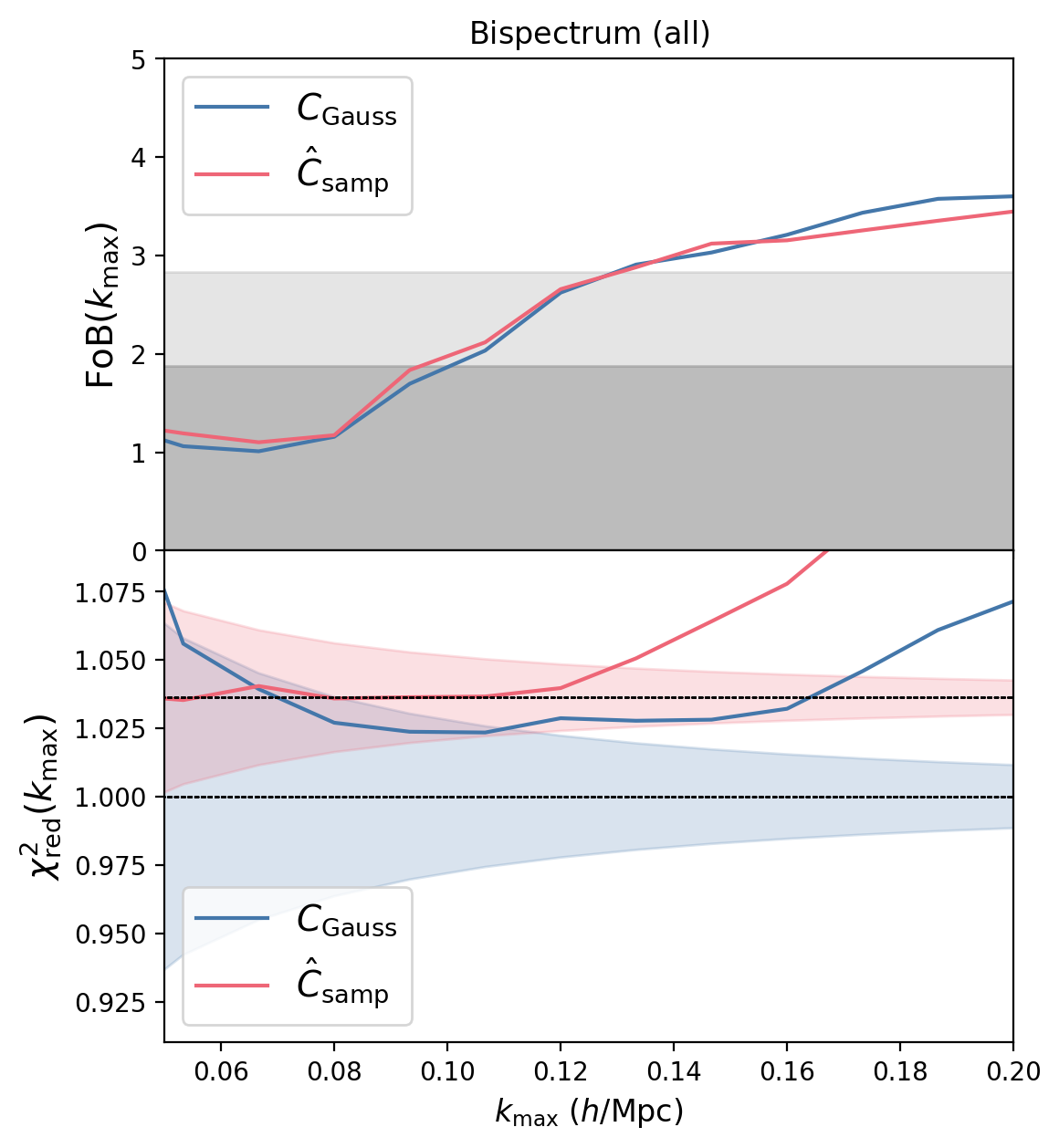}
    \caption{FoB and $\chi_\text{red}^2$ for the Gaussian (blue) and diagonal sample covariance (pink) for the full likelihood from Eq. \eqref{eq:liks} (cf. Figure \ref{fig:fitplot_sst}).}
    \label{fig:allfitplot}
\end{figure}
\begin{equation}\label{eq:stochamp}
    \alpha_\text{gg}^{1\delta} = 4.56 \pm 0.49. 
\end{equation}
The best-fit value is in excellent agreement with the Poissonian expectation from Eq. \eqref{eq:pois}; note however that in the EFT framework, noise amplitudes are treated as free parameters and can in principle deviate by $\mathcal{O}(1)$ from their Poisson values. For example, in \cite{bakx_eft} it was found that the shape noise amplitude $c^\text{g}$ from Eq. \eqref{eq:nlamod} differs from its Poisson expectation by $\sim 5\%$ \cite{kurita_ia} (and this difference, although small, was statistically significant). One feature that distinguishes IA as a tracer of the dark matter field is that the IA field is much noisier than e.g. halo (or galaxy) number counts. Thus, relative contributions of stochastic terms to the total spectrum can be large. For $B_{\delta EE}$ the amplitude of the noise term is numerically comparable to the deterministic part. This is no longer the case for $B_{\delta EB}$. Indeed, we have checked that already on large scales $k \sim 0.05\,h/\text{Mpc}$ this term nominally comprises $\sim 50\% (80\%)$ of the signal for $B_{\delta EE}\, (B_{\delta EB})$.

Figure \ref{fig:fitplot_stt} shows the performance metrics from the previous Section. We again see strong agreement between the FoB for both choices of covariance; however this time $\chi_\text{red}^2$ is somewhat higher for the diagonal sample variance. If one would trust only the most pessimistic estimate and use $\Delta_{\text{crit}} = 1.5$, the maximum allowed scale would decrease to $k_\text{max} \approx 0.09\,h/\text{Mpc}$, inflating posteriors on bias parameters by nominally $\sim 50\%$. As a middle ground, if one either switches to $\Delta_{\text{crit}} = 2$ or uses the Gaussian covariance with $\Delta_{\text{crit}} = 1.5$, the previously found scale cut of $k_\text{max} \approx 0.11\,h/\text{Mpc}$ is reasonable. Moreover, we will see that this scale cut remains reasonable also when adding bispectra with three tensor fields; we therefore continue quoting it as our fiducial choice. The differences in $\chi_\text{red}^2$ also lead to similar differences in the breakdown criterion $\Delta(k_\text{max})$ in Figure \ref{fig:deltaall} (pink lines). 

Finally, as a means of checking the compatibility with the parity-odd multipole $B_{\delta EB}^{02}$ again we also show in Figure \ref{fig:sigplot_stt} the bispectrum signal plotted for the set of parameters listed in Eq. \eqref{eq:plotparams} (that is, the same set that was used for the bispectra with one tensor field) along with the additional choice of $\alpha_\text{gg}^{1\delta} = 4.5$ (cf. Eq. \eqref{eq:gg1d}). These bispectra are generally noisier than the ones with only one shape field, so the agreement is visually better in most cases. However, the amplitude of $B_{\delta B B}^{02}$ prefers lower values of $\alpha_\text{gg}^{1\delta}$ than those preferred by $B_{\delta EE}^{02}$, as exemplified by the middle left plot. This is simply due to the latter dominating the overall SNR. Note that this time, the parity-odd bispectrum $B_{\delta EB}$ does have a strong signal in the equilateral configuration (see also Figure 4 in Paper I); it no longer vanishes because of symmetries since all three fields are distinct. 
\subsection{One, two or three shape fields}\label{subsec:threefield}
For bispectra with three tensor fields, we have four nontrivial combinations, to wit $B_{EEE}, B_{EEB}, B_{EBB},B_{BBB}$. We again measure their lowest order multipoles, that is $(\ell_1,\ell_2) = (2,2)$. We add the parity even bispectra $\tilde{B}^{22}_{EEE}, \tilde{B}^{22}_{EBB}$ to the baseline likelihood\footnote{When using the diagonal sample covariance estimator, we again set the cross-covariance between $B_{E BB}$ and all multipoles not containing B-modes in the same triangle bin to zero; this is what is expected in the Gaussian approximation. We do however include the cross-covariance between $\tilde{B}^{02}_{\delta BB}$ and $\tilde{B}^{22}_{EBB}$, which is nonzero. Thus, in a given triangle bin the estimator from Eq. \eqref{eq:covest} consists of a 5 by 5 block and a 2 by 2 block.}. Similarly to $B_{\delta BB}$, the $B_{EBB}$ bispectrum depends only on the noise amplitude. While $\tilde{B}^{22}_{EEE}$ also depends on all the first and second order operator biases, it is completely dominated by the contribution proportional to $\alpha_\text{gg}^{1\delta}$, contrary to $\tilde{B}^{02}_{\delta EE}$. 

When performing the same exercise as in the previous two cases, we again find good agreement between the data and the tree-level EFT of IA at $k_\text{max} = 0.11\,h/\text{Mpc}$, as shown by the red contour in Figure \ref{fig:corner_all}.
The same stochastic amplitude $\alpha_\text{gg}^{1\delta}$ that previously described the two-tensor bispectra also provides a good description of the three-tensor bispectra. This is a rather nontrivial prediction of the EFT of IA which is strongly supported by the data. We obtain 
\begin{equation}\label{eq:gg1d}
    \alpha_\text{gg}^{1\delta} = 4.50 \pm 0.38;
\end{equation}
in good agreement with Eq. \eqref{eq:stochamp} and the Poisson prediction from Eq. \eqref{eq:pois}.

Figure \ref{fig:allfitplot} shows the same performance metrics again for all seven parity-even multipoles combined. This time, the choice of Gaussian covariance seems to yield an earlier breakdown of the model due to a consistently higher $\chi_\text{red}^2$ (solid green line in Figure \ref{fig:deltaall}); however, we emphasize again that this statistic is very sensitive to the correctness of the covariance, and so needs to be interpreted with care. The FoB exceeds its $2\sigma$ threshold at roughly $k_\text{max} = 0.12\,h/\text{Mpc}$, as it does in Figure \ref{fig:fitplot_sst} and Figure \ref{fig:fitplot_stt} - we regard this as the most robust indicator that the model is failing (though see the discussion below Eq. \eqref{eq:b21res}). Note that when Eq. \eqref{eq:covest} is used for the covariance, the pink line in the bottom panel of Figure \ref{fig:allfitplot} closely follows the correct theoretical mean $\hat{\mu}_\text{red}$ from Eq. \eqref{eq:chisqresult}, which is different from unity.

Moreover, one can also see from Figure \ref{fig:snreven} that the B-mode bispectrum $B^{22}_{BBB}$, which is zero in our theory model, is not detected at any significance ($\text{SNR}<3$) up to $k \approx 0.11\,h/\text{Mpc}$. Beyond this wavenumber, gradient corrections to stochastic terms (or, albeit somewhat less likely, loops from the bias expansion) become important, which can have nontrivial projection onto the product of three B-modes. This serves as an additional confirmation for our choice of scale cut. 

\begin{figure}[t]
    \centering
    \includegraphics[width=\linewidth]{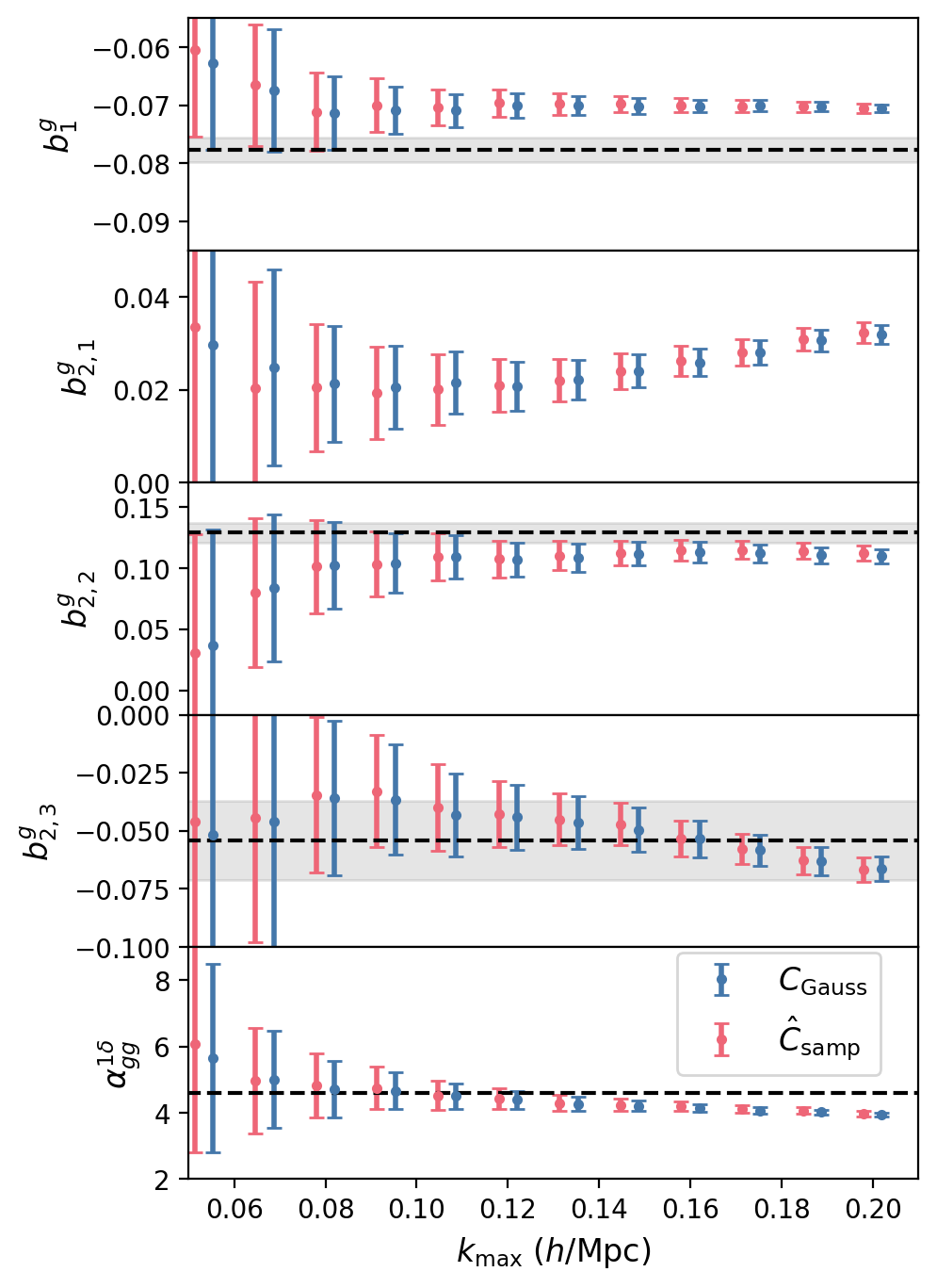}
    \caption{Dependence of the five fitting parameters from Eq. \eqref{eq:modelpars} on the maximum scale included, either for Gaussian covariance (blue) or Eq. \eqref{eq:covest} (pink). For $b_1^\text{g}, b_{2,2}^\text{g}$ and $b_{2,3}^\text{g}$ the shaded bands indicate the uncertainties from Eqs. \eqref{eq:b1fid} and \eqref{eq:b2fid}. In the bottom panel, the dashed line indicates the Poisson expectation from Eq. \eqref{eq:pois}. We do not plot an error band for $b_{2,1}^\text{g}$ since the power spectrum posterior for this parameter is bimodal.}
    \label{fig:runplot}
\end{figure}

In addition, it is important to verify that bias parameters do not run with the scales included in the fit. Figure \ref{fig:runplot} provides an illustration of this statement in the case where all multipoles are combined (but we also performed this check for all fits separately). We see that for $k_\text{max} \gtrsim 0.14\,h/\text{Mpc}$, some bias parameters start to drift away from their values at lower $k_\text{max}$ (most notably $b_{2,1}^\text{g}$ and $b_{2,3}^\text{g}$), which implies that the model tries to absorb higher-order effects we did not include in the fit and thus is no longer adequate.

Previously, the power spectrum-only analysis of \cite{bakx_eft} was complicated to some extent by the quadratic dependence of the model predictions on the second order bias parameters. In that work, parameter posteriors of the second order bias parameters $b_{2,1}^\text{g}$ and $b_{2,3}^\text{g}$ were found to exhibit bimodalities when analyzing only the one-loop power spectrum of IA. To mitigate this issue, \cite{bakx_eft} assigned a physically motivated prior on the signs of these parameters as expected from a local Lagrangian bias assumption \cite{akitsu_bias,hymalaia}. We showed that this assumption is correct and that the degeneracies present in the one-loop power spectrum analysis can be broken through the addition of the bispectrum. Indeed, in the tree-level bispectrum, all second order bias parameters appear linearly in the model (in fact, in the case of e.g. $B_{\delta \delta E}$ there is only one biased tracer, so all parameters appear linearly in the model). Thus, the degeneracies present in the power spectrum likelihood are guaranteed to be broken. 
Our conclusive determination of the second order bias parameters has some immediate interesting phenomenological applications. For example, in terms of the more familiar second order basis $\{b_{\delta K}, b_{K^2}, b_t\}$ (see e.g. \cite{bakx_eft} for definitions of these operators and the corresponding basis transformation) we obtain the parameter constraints
\begin{equation}
\begin{aligned}
    b_{\delta K} &= \frac{2}{3}b_{2,1}^\text{g} +\frac{1}{3} b_{2,3}^\text{g} = 0.000 \pm 0.007; \\
    b_{K^2} &= b_{2,1}^\text{g} - b_{2,3}^\text{g} = 0.060 \pm 0.021; \\
    b_t &= \frac{7}{8}(b_{2,3}^\text{g}-b_{2,2}^\text{g}) = -0.131 \pm 0.021;
\end{aligned}
\end{equation}
thus demonstrating good evidence of positive tidal torquing and a strong detection of the velocity shear operator $t_{ij}$. This operator may contain interesting information about the formation history of the halo, since it depends non-locally (in time) on the dark matter density field, as discussed in \cite{schmitz_bisp}. More precisely, if galaxies acquire their intrinsic alignment through instantaneous response to the local tidal field at some redshift $z_*$, the contribution of $t_{ij}$ to the bias expansion is generated entirely through advection and can thus be calculated given $z_*$ (and additional cosmological parameters that determine the statistics of the dark matter density field).  

Lastly, we again plot the measured multipole signal for bispectra with three tensor fields against the theoretical predictions evaluated with Eq. \eqref{eq:plotparams} and $\alpha_\text{gg}^{1\delta} = 4$ as for the case with two tensor fields. The result is shown in Figure \ref{fig:sigplot_ttt}. We again observe good agreement even for the larger volume of $V_\text{DESI} = 7.5\,(\text{Gpc}/h)^3$. Note that the signal for $B_{BBB}^{22}$ (bottom panel) is very weak, in line with our tree-level expectation.

\section{Conclusions and Discussion}\label{sec:discussion}
We have seen that the tree-level bispectrum in the EFT of IA accurately captures the rich scale- and angle dependence of the bispectrum of IA. In particular, a five-parameter model involving four deterministic bias parameters occurring in the second order `TATT' expansion and an additional noise term provides a good fit to the different bispectra considered in Section \ref{sec:results} up to $k_\text{max}=0.11\,h/\text{Mpc}$. Moreover, the parameters that are shared with the power spectrum of matter and shape fields match those from the power spectrum analysis, and the noise parameter is close to its Poisson expectation. This is a nontrivial achievement, given that the cumulative SNR of all bispectra in a single volume of $V=1\,(\text{Gpc}/h)^3$ is around $\approx 30$ (cf. Figure \ref{fig:snreven}) for $k_\text{max} = 0.11\,h/\text{Mpc}$.

Note that simpler models for the scale dependence of intrinsic alignment bispectra fall short here; for example according to NLA-inspired models \cite{pyne_3pt, linke_lowz} there is no B-mode and we would have $B_{\delta \delta B}=0$. However, we have clearly demonstrated that this bispectrum is nonzero. In addition, our work indicates that bispectra with more than one shape field are strongly dependent on the noise amplitude $\alpha_{\text{gg}}^{1\delta}$. This contribution was missing in the earlier second-order (i.e. `TATT') alignment models of \cite{blazek_tatt,schmitz_bisp}. We emphasize that the importance of this contribution will only be exacerbated as we consider more realistic samples with lower number densities and lower alignment amplitudes. To this end, it is reassuring that several different bispectra can in principle be combined in order to jointly constrain noise amplitudes. This is also important when attempting to use the bispectrum of IA for mitigation purposes in cosmic shear surveys. Generally speaking, other models for IA which are inspired by the galaxy bias expansion are also capable of providing physically motivated templates for the IA bispectrum that can be used in conjunction with two-point statistics. This may allow for more efficient nuisance parameter choices \cite{hymalaia}. Alternatively, using the Lagrangian framework, \cite{chen_lpt,chen_lensing} showed that a perturbative description of intrinsic alignments can be used to place conservative constraints on the amplitude of matter fluctuations $S_8 \equiv \sigma_8(\Omega_m/0.3)^{0.5}$.We anticipate that the bispectrum of IA can be a useful tool in directly constraining the intrinsic alignment of galaxy samples when spectroscopic redshifts are available \cite{samuroff_ia,hervaspeters}.

Moreover, we have seen that higher multipoles are instrumental in breaking degeneracies amongst second order bias parameters (cf. Figure \ref{fig:degenplot}). This could not be established by only considering the SNR of these multipoles, as we did in Paper I. The different multipoles thus provide an efficient way of incorporating information from the full, anisotropic bispectrum $B_{\delta \delta E}$. We also found that when adding bispectra with two shape fields, constraints on bias parameters improved significantly (cf. Figure \ref{fig:corner_all}). However, this conclusion may depend strongly on the aforementioned sample properties such as alignment amplitude and shot noise. 

Our results motivate a more in-depth study of joint constraints wrought from the power spectrum and bispectrum of IA for a more realistic halo or galaxy sample. This can be done by downsampling halos from an N-body simulation and randomizing their orientations to mimic misalignment of the LRG with respect to its host \cite{okumura_misalignment,vanalfen}, or by considering a galaxy population from a large-scale hydrodynamical simulation such as \cite{flamingo}. We will explore these options in future work. 
\begin{acknowledgments}
We thank Fabian Schmidt for useful discussions.

This publication is part of the project ``A rising tide: Galaxy intrinsic alignments as a new probe of cosmology and galaxy evolution'' (with project number VI.Vidi.203.011) of the Talent programme Vidi which is (partly) financed by the Dutch Research Council (NWO). Z.V. acknowledges the support of the Kavli Foundation. 
\end{acknowledgments}
\newpage
\appendix
\section{A: Theory Expressions for Tree-level Tensor Bispectrum}\label{sec:expr}
The tree-level tensor bispectrum contains a deterministic and a stochastic piece. 
The $n$-th order term in Fourier space for the deterministic piece takes the form
\begin{equation}
    S_{ij}^{(n)}(\mathbf{k}) = \int \md^3\bb{p}_1 \dots \md^3\bb{p}_n\, \mathcal{K}_{ij}^{(n)}(\mathbf{p}_1, \dots \mathbf{p}_n) \delta^D(\mathbf{k}-\mathbf{p}_{1,n}) \delta^{(1)}(\mathbf{p}_1) \dots \delta^{(1)}(\mathbf{p}_n)
\end{equation}
where
\begin{equation}
    \mathcal{K}_{ij}^{(n)} = \mathcal{K}_{ij}^{\text{s},(n)}+\mathcal{K}_{ij}^{\text{g},(n)}
\end{equation}
is the decomposition into trace- and trace-free parts and $\delta^{(1)}$ is the linear matter density field. The total deterministic bispectrum at tree-level considered in this work is thus
\begin{equation}\label{bispec}
\begin{aligned}
B_{ijklrs}^{\alpha \beta \gamma, \text{det,tree}}(\mathbf{k}_1,\mathbf{k}_2,\mathbf{k}_3) 
&= 2\,\mathcal{K}_{ij}^{\alpha,(1)}(\mathbf{k}_1)\mathcal{K}_{kl}^{\beta,(1)}(\mathbf{k}_2)\mathcal{K}_{rs}^{\gamma,(2)}(\mathbf{k}_1,\mathbf{k}_2)P_L(k_1)P_L(k_2)+ \text{2 perm.} 
\end{aligned}    
\end{equation}
where $\alpha,\beta,\gamma$ are either $\delta$ or g. Since we only consider the matter density field in the scalar sector, the scalar kernels are trivial. Explicitly, the first order kernels are given by
\begin{equation}
\begin{aligned}
    \mathcal{K}_{ij}^{\delta,(1)}(\mathbf{k}) &= \frac{1}{3}\delta_{ij}^\mathrm{K}; \\
    \mathcal{K}_{ij}^{\text{g},(1)}(\mathbf{k}) &= b_1^\text{g} \big( \frac{\bb{k}_i\bb{k}_j}{k^2}-\frac{1}{3}\delta_{ij}^\mathrm{K}\big) = b_1^\text{g}\mathcal{D}_{ij}(\bb{k});
\end{aligned}
\end{equation}
where in the second line we introduced the trace-free Fourier space operator $\mathcal{D}_{ij}(\bb{k}) = \bb{k}_i\bb{k}_j/k^2-\frac{1}{3}\delta_{ij}^\mathrm{K} $.
The second order kernels are given by
\begin{equation}\label{eq:secker}
    \begin{aligned}
        \mathcal{K}_{ij}^{\delta,(2)}(\mathbf{k}_1,\bb{k}_2) &= \frac{1}{3}\delta_{ij}^\mathrm{K}F_2(\mathbf{k}_1,\bb{k}_2); \\
        \mathcal{K}_{ij}^{\text{g},(2)}(\mathbf{k}_1,\bb{k}_2) &= \text{TF}\bigg[c_{1}^\text{g} \bigg(\frac{\bb{k}_{12,i}\bb{k}_{12,j}}{k_{12}^2}F_2(\mathbf{k}_1,\bb{k}_2)\bigg) + c_{2,1}^\text{g}\bigg(\frac{\bb{k}_{12,i}\bb{k}_{12,j}}{k_{12}^2}F_2(\mathbf{k}_1,\bb{k}_2)-\frac{1}{2}\frac{\bb{k}_1\cdot \bb{k}_2}{k_1^2k_2^2}(\bb{k}_{1,i}\bb{k}_{1,j}+\bb{k}_{2,i}\bb{k}_{2,j})\bigg) \\
        &+ c_{2,2}^\text{g}\bigg(\frac{1}{2}\frac{\bb{k}_1\cdot \bb{k}_2}{k_1^2k_2^2}(\bb{k}_{1,i}\bb{k}_{2,j}+\bb{k}_{2,i}\bb{k}_{1,j})\bigg) + c_{2,3}^\text{g} \bigg( \frac{\bb{k}_{1,i}\bb{k}_{1,j}}{2k_1^2}+\frac{\bb{k}_{2,i}\bb{k}_{2,j}}{2k_2^2}\bigg)\bigg]
        \end{aligned}
\end{equation}
where $\bb{k}_{12} = \bb{k}_1+\bb{k}_2$ and $F_2$ is the usual the second order scalar kernel.

In the notation of \cite{mirbabayi}, the coefficients $c_1^{\text{g}},c_{2,a}^{\text{g}} (a=1,2,3)$ belong to the trace-free parts of the operators
\begin{equation}
    \Pi^{[1]}_{ij}, \Pi^{[2]}_{ij},(\Pi^{[1]})^2_{ij}, \Pi^{[1]}_{ij}\text{Tr}(\Pi^{[1]}_{ij})
\end{equation} 
respectively. 
We work with a redefinition of shape bias coefficients which is given by 
\begin{equation}
\begin{aligned}
    b_1^\text{g} &=c_1^\text{g}; \qquad b_{2,1}^\text{g} = c_{2,1}^\text{g}+c_{2,2}^\text{g}+c_{2,3}^\text{g}; \qquad b_{2,2}^\text{g} = c_{2,3}^\text{g}+\frac{20}{7}c_{2,1}^\text{g}; 
    \qquad b_{2,3}^\text{g}=c_{2,3}^\text{g}. 
\end{aligned}
\end{equation}
The B-mode does not depend on $b_{2,2}^\text{g}$, which automatically implies that bispectra containing any B-mode do not depend on $b_{2,2}^\text{g}$ at tree level (since the B-mode starts at second order). We also use the relation to another basis used consisting of the operators $b_{\delta K}, b_{K^2}$ and $b_t$, which is 
\begin{equation}
    b_{\delta K} = \frac{1}{3}(2b_{2,1}^\text{g}+b_{2,3}^\text{g}), \quad b_{K^2} = b_{2,1}^\text{g} - b_{2,3}^\text{g}, \quad b_t = \frac{7}{8}(b^\text{g}_{2,3}-b^\text{g}_{2,2}).
\end{equation}
We will need the co-evolution relations for the second order parameters in terms of $b_1^\text{s}$ and $b_1^\text{g}$, which are derived under the assumption that the Lagrangian higher order bias parameters are zero. They read 
\begin{equation}
    b_{2,1}^\text{g} = (b_1^\text{s}-1)b_1^\text{g}; \qquad b_{2,2}^\text{g} = (b_1^\text{s}-20/7)b_1^\text{g}; \qquad b_{2,3}^\text{g} = b_1^\text{s}b_1^\text{g}.
\end{equation}
For the stochastic piece, all the relevant correlators were collected in Paper I. The only nonzero stochastic contribution we can form with the stochastic shape contributions 
\begin{equation}
    g_{ij}^{\text{stoch}}(\bb{x}) \qquad \supset \qquad \epsilon_{ij}(\mathbf{x}) \text{ (first order)}; \qquad \epsilon_{ij}^\delta(\bb{x})\delta(\bb{x}),\quad \epsilon^\Pi(\mathbf{x})\text{TF}(\Pi^{[1]})_{ij}(\bb{x}) \text{ (second order) }
\end{equation}
and which occurs in the projected tree-level bispectrum is
\begin{equation}
    \langle \epsilon_{ij}(\mathbf{k}_1)\epsilon_{kl}^\delta(\mathbf{k}_2)\rangle ' = \alpha_{\text{gg}}^{1\delta}\Delta_{ij,kl}^{\text{TF}}  + \mathcal{O}((k/k_{\text{NL}})^2);
\end{equation}
where 
\begin{equation}
    \Delta_{ij,kl}^{\text{TF}} = \delta_{ik}^\mathrm{K}\delta_{jl}^\mathrm{K}+\delta_{il}^\mathrm{K}\delta_{jk}^\mathrm{K}-\frac{2}{3}\delta_{ij}^\mathrm{K}\delta_{kl}^\mathrm{K}
\end{equation}
is pair-symmetric and pair-traceless.
This combination occurs in spectra with two or three tensor fields, specifically as 
\begin{equation}
    \begin{aligned}
        B_{ijkl}^{\delta\text{gg}} &\supset \langle \delta(\bb{k}_1)\epsilon_{ij}(\mathbf{k}_2)(\epsilon_{kl}^\delta\delta)(\mathbf{k}_3)\rangle ' =  \alpha_{\text{gg}}^{1\delta} \Delta_{ij,kl}^{\text{TF}}P_L(k_1) + \mathcal{O}((k_i/k_{\text{NL}})^2);\\
    B_{ijklrs}^{\text{ggg}} &\supset \langle (b_1^\text{g}\text{TF}(\Pi^{[1]})_{ij})(\bb{k}_1)(\epsilon_{kl}^\delta\delta)(\mathbf{k}_2)\epsilon_{rs}(\mathbf{k}_3)\rangle ' = b_1^\text{g}\alpha_{\text{gg}}^{1\delta} \mathcal{D}_{ij}(\bb{k}_1) \Delta_{kl,rs}^{\text{TF}}P_L(k_1)+ \mathcal{O}((k_i/k_{\text{NL}})^2). 
    \end{aligned}
\end{equation}
For brevity, we resort to reproducing the table containing all dependencies of the spectra on respective bias parameters (see Table \ref{tab:params}).
\begin{table}
\begin{center}
\begin{tabular}{|c|c|c|c|c|c|c|}
\hline
    $B$ / Bias & $b_1^{\text{g}}$ & $b_{2,1}^{\text{g}}$ & $b_{2,2}^{\text{g}}$ & $b_{2,3}^{\text{g}}$ &  $\alpha_\text{gg}^{1\delta}$ \\
    \hline
    $B_{\delta \delta E}$  & \cm & \cm & \cm & \cm & $\crs$  \\
    \hline
    $B_{\delta \delta B}$ & $\crs$ & \cm & $\crs$ & \cm & $\crs$ \\
    \hline
    $B_{\delta E E}$ & \cm & \cm & \cm & \cm & \cm    \\
    \hline
    $B_{\delta B B}$ & $\crs$ & $\crs$ & $\crs$ & $\crs$ & \cm   \\
    \hline
    $B_{\delta E B}$ & \cm & \cm & $\crs$ & \cm &  \cm   \\
    \hline
    $B_{E E E}$ & \cm & \cm & \cm & \cm & \cm    \\
    \hline
    $B_{E E B}$ & \cm & \cm &  $\crs$ & \cm  & \cm \\
    \hline
    $B_{E B B}$ & $\crs$ & $\crs$ & $\crs$ & $\crs$ &\cm  \\
    \hline
\end{tabular}
\caption{Parameter dependence of tensorial bispectra.\label{tab:params}}
\end{center}
\end{table} 
\section{B: Details on Covariance and Model Fits}\label{sec:cov}
\subsection{Further comparison between Gaussian and sample covariance}
\begin{figure}
    \centering
    \includegraphics[width=0.5\linewidth]{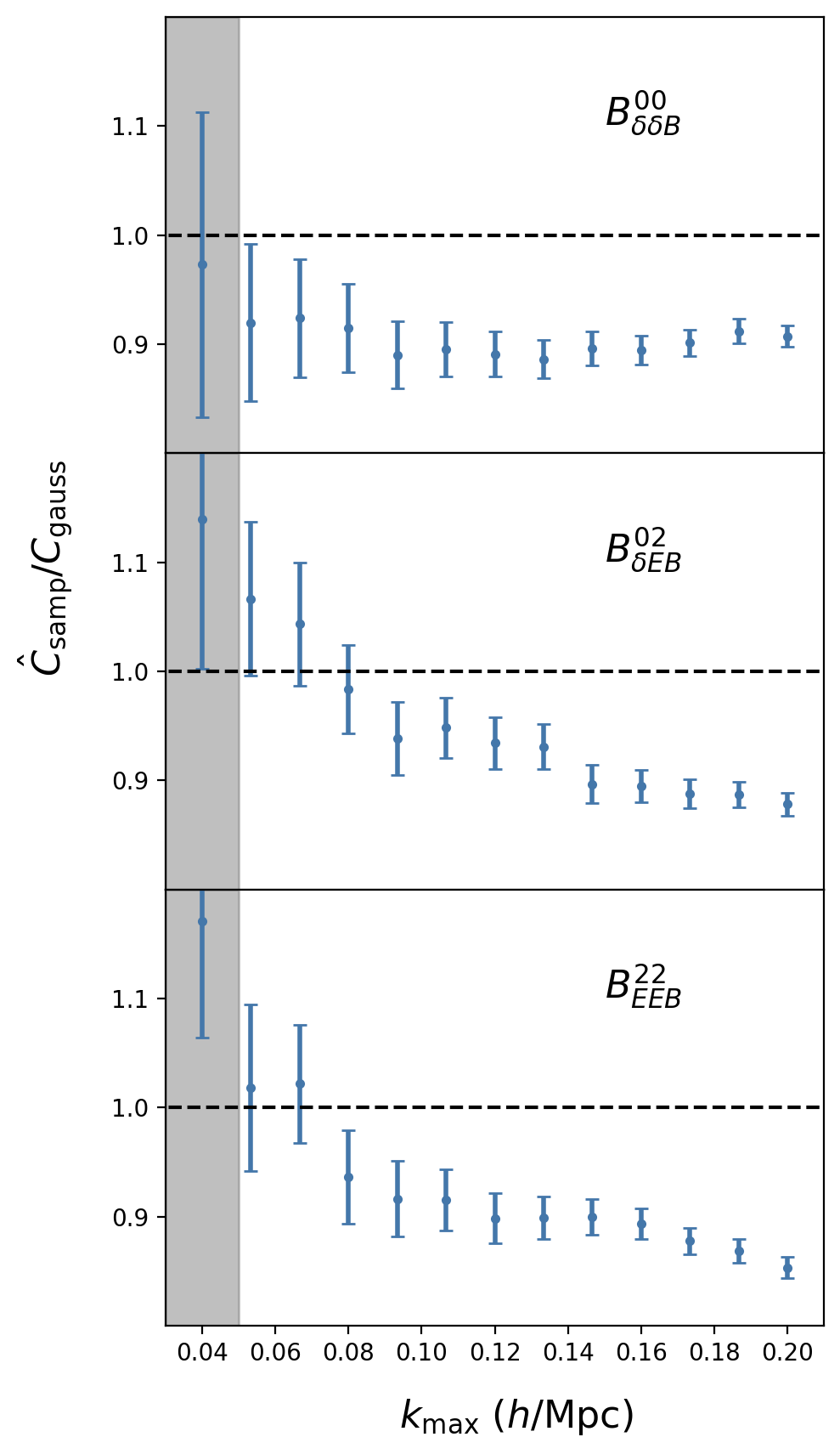}
    \caption{Ratio of estimated and Gaussian variance for the three parity-odd 
    bispectra for which we predict a nonvanishing signal. The grey shade indicates the region where $k_\text{max} < 0.05\,h/\text{Mpc}$; deviations in this region are likely due to discreteness effects and do not indicate model failure.}
    \label{fig:cov_comp_odd}
\end{figure}

Here we list some further tests comparing the sample covariance from Eq. \eqref{eq:covest} to the theoretical Gaussian one. In particular, we plot the same ratio as that of Figure \ref{fig:cov_comp} for the variance of the four odd bispectra. The result is shown in Figure \ref{fig:cov_comp_odd}. We again broadly find agreement between the two methods, with more pronounced differences for higher values of $k_\text{max}$. 
The Gaussian covariance tends to overestimate the errors at smaller scales. 
\subsection{Undoing chi-squared bias when using the diagonal sample covariance}
Using a noisy estimate for the covariance in a Gaussian likelihood can lead to underestimated errors on parameters if not properly taken into account. In particular, following \cite{hartlap} we see that in order to construct an unbiased estimator of the true log-likelihood
\begin{equation}
    \chi^2_{\text{true}} = -2 \log \mathcal{L} =  \sum_{L,L'}^{k_\text{max}}(\hat{\bar{B}}_L-B_L^\text{th})(C^\text{true}_{LL'})^{-1}(\hat{\bar{B}}_{L'}-B_{L'}^\text{th})
\end{equation}
we cannot use the inverse of the diagonal sample variance of Eq. \eqref{eq:covest}. Indeed, 
\begin{equation}\label{eq:chisqexp}
    \bigg\langle  \sum_{L,L'}^{k_\text{max}}(\hat{\bar{B}}_L-B_L^\text{th})(\hat{C}_{LL'})^{-1}(\hat{\bar{B}}_{L'}-B_{L'}^\text{th}) \bigg\rangle = \sum_{L,L'}^{k_\text{max}} \langle (\hat{\bar{B}}_L-B_L^\text{th})(\hat{\bar{B}}_{L'}-B_{L'}^\text{th}) \rangle \langle (\hat{C}_{LL'})^{-1} \rangle = \frac{N_r - 1}{N_r - N_L - 2}\langle \chi^2_{\text{true}} \rangle 
\end{equation}
where in the second equality we used the independence of the sample mean and sample variance to factor the expectation value \cite{anderson2003} and $N_L$ is the number of multipoles. The Hartlap prefactor needs to be applied to every block in multipole space in order to get an unbiased estimate of the true precision matrix, where the size of the sample covariance is $N_L \times N_L$ (assuming that the true covariance is diagonal in triangle space). This factor is greater than unity, indicating that error bars will be underestimated if this correction is not applied. In the limit where $N_r \gg N_L$, this prefactor approaches 1 as expected. 

Naively, one might expect that a similar correction should be applied to the goodness-of-fit statistic, that is, to obtain an unbiased estimator of 
\begin{equation}
    \bar{\chi}_\text{true}^2 = \frac{1}{N_r} \sum_{L,L',r}^{k_\text{max}}(\hat{B}^r_{L}-B_L^\text{th})(C^\text{true}_{LL'})^{-1}(\hat{B}^r_{L'}-B_{L'}^\text{th}), 
\end{equation}
we should multiply the `naive' expression (where the hat indicates that we are using a sample covariance) 
\begin{equation}\label{eq:chibarest}
    \hat{\bar{\chi}}^2 = \frac{1}{N_r} \sum_{L,L',r}^{k_\text{max}}(\hat{B}^r_{L}-B_L^\text{th})(\hat{C}_{LL'})^{-1}(\hat{B}^{r}_{L'}-B_{L'}^\text{th})
\end{equation}
by the reciprocal of Eq. \eqref{eq:hartlapfac}. However, this intuition is not correct. The reason is that while the sample mean $\hat{\bar{B}}_L$ and the sample covariance $\hat{C}_{LL'}$ are independent, the same is not true for the individual realizations $\hat{B}^r_{L}$ and the expectation does not factorize (this property was used in \cite{hartlap}). However, we can still compute the expectation value of Eq. \eqref{eq:chibarest} (when the covariance is taken to be the uncorrected block diagonal sample variance from Eq. \eqref{eq:covest}). To wit, after some algebra we obtain
\begin{equation}\label{eq:chisqid}
    \hat{\bar{\chi}}^2 = \hat{\chi}^2 + \frac{1}{N_r}\sum_{L,L',r}^{k_\text{max}}(\hat{B}^r_L-\hat{\bar{B}}_{L})(\hat{C}_{LL'})^{-1}(\hat{B}^{r}_{L'}-\hat{\bar{B}}_{L'}).
\end{equation}
without taking the expectation value, where 
\begin{equation}
    \hat{\chi}^2 =  \sum_{L,L'}^{k_\text{max}}(\hat{\bar{B}}_L-B_L^\text{th})(\hat{C}_{LL'})^{-1}(\hat{\bar{B}}_{L'}-B_{L'}^\text{th}).
\end{equation}
Note that Eq. \eqref{eq:chisqid} is true regardless of the structure of the covariance matrix in question. However, when $\hat{C}$ is the sample covariance, the second term in Eq. \eqref{eq:chisqid} is \textit{independent of the data} and simply reads 
\begin{equation}
\begin{aligned}
    \frac{1}{N_r}\sum_{L,L',r}^{k_\text{max}}(\hat{B}^r_L-\hat{\bar{B}}_{L})(\hat{C}_{LL'})^{-1}(\hat{B}^r_{L'}-\hat{\bar{B}}_{L'}) &= \frac{1}{N_r}\sum_{L,L'}^{k_\text{max}}(\hat{C}_{LL'})^{-1}\sum_r (\hat{B}^r_L-\hat{\bar{B}}_{L})(\hat{B}^r_{L'}-\hat{\bar{B}}_{L'}) \\
    &= \frac{N_r-1}{N_r}\sum_{L,L'}^{k_\text{max}}(\hat{C}_{LL'})^{-1} \hat{C}_{LL'} \\
    &= \frac{N_T N_L(N_r-1)}{N_r}
\end{aligned}
\end{equation}
after plugging in Eq. \eqref{eq:covest}. Thus, after taking the expectation value of Eq. \eqref{eq:chisqid} and using Eq. \eqref{eq:chisqexp} we obtain 
\begin{equation}
    \langle \hat{\bar{\chi}}^2 \rangle =\langle \hat{\chi}^2 \rangle  + \frac{N_T N_L(N_r-1)}{N_r} = \frac{N_r-1}{N_r}\bigg(\frac{N_r \langle \chi_\text{true}^2 \rangle }{N_r - N_L - 2} + N_T N_L \bigg).
\end{equation}
Similarly, in case of using true covariance $C^\text{true}$, the analog of Eq. \eqref{eq:chisqid} still holds, and after taking expectation values on both sides and using $\langle \bar{\chi}_\text{true}^2 \rangle = N_TN_L$ we obtain 
\begin{equation}
    \langle \bar{\chi}_\text{true}^2 \rangle = \langle \chi_\text{true}^2 \rangle  + \frac{N_T N_L(N_r-1)}{N_r} \implies N_r \langle \chi_\text{true}^2 \rangle = N_r \langle \bar{\chi}_\text{true}^2 \rangle - N_T N_L (N_r - 1) = N_T N_L.
\end{equation}
Finally, after inserting the above equality into the expression for $\langle \hat{\bar{\chi}}^2 \rangle$ and dividing by the number of degrees of freedom $N_T N_L$ we infer that
\begin{equation}\label{eq:chisqresult}
    \hat{\mu}_\text{red} \equiv \langle \hat{\bar{\chi}}_\text{red}^2 \rangle = \frac{\langle \hat{\bar{\chi}}^2 \rangle}{N_T N_L} = \frac{N_r - 1}{N_r}\bigg( 1 + \frac{1}{N_r - N_L -2}\bigg),
\end{equation}
where the hat on $\hat{\mu}_\text{red}$ indicates that the block diagonal sample covariance is used in $\hat{\chi}^2$. Typically, when $N_r \gg N_L $ and $N_r \gg 1$ the difference between Eq. \eqref{eq:chisqresult} and unity is $\mathcal{O}(N_L/N_r^2) \ll 1$. However, this correction can be at the percent level when e.g. $N_r = 20$ and $N_L = 5$. More importantly, the variance of $\hat{\bar{\chi}}^2$ is\footnote{To be more precise, the correct distribution of $\hat{\chi}^2$ is \cite{hotelling1931generalization} $$N_r\hat{\chi}^2 \sim \sum_{n=1}^{N_T}X_n$$ where all the $X_n \sim  T^2_{N_L,N_r-1}$ are independent and $T^2$-distributed. As such, the variance of the sum is given by (see e.g. \cite{johnson1995continuous}) $$ \text{Var}(N_r\hat{\chi}^2) = N_T \text{Var}(X_n) = \frac{2N_T N_L(N_r-1)^2(N_r-2)}{(N_r-N_L-2)^2(N_r-N_L-4)}.$$} 
\begin{equation}
    \text{Var}(\hat{\bar{\chi}}^2) = \text{Var}(\hat{\chi}^2) = \frac{2 N_T N_L (N_r - 1)^2(N_r-2)}{N_r^2(N_r - N_L - 2)^2(N_r-N_L-4)}.
\end{equation}
and finally 
\begin{equation}\label{eq:varred}
    \hat{\sigma}_\text{red} \equiv (\text{Var}(\hat{\chi}_\text{red}^2 ))^{1/2} = \bigg(\text{Var}\bigg(\frac{\hat{\bar{\chi}}_*^2}{N_TN_L}\bigg)\bigg)^{1/2}  = \bigg(\frac{2  (N_r - 1)^2(N_r-2)}{N_r^2N_T N_L(N_r - N_L - 2)^2(N_r-N_L-4)}\bigg)^{1/2}.
\end{equation}
Crucially, when $N_r \to \infty$ this result does \textit{not} exhibit the naively expected behaviour of $\sigma_\text{red} \sim \sqrt{2/(N_T N_L N_r)}$ based on counting the number of degrees of freedom, but rather
\begin{equation}
    \sigma_\text{red} \sim \sqrt{2/(N_T N_L N_r^{\boldsymbol{2}})}.
\end{equation}
That is, the scatter in the reduced chi-squared computed \textit{from summing all the individual realizations} is \textit{smaller} when using the sample covariance than when using a noiseless covariance. To our knowledge, this result has not been appreciated in the literature thus far. On the other hand, if one does not sum the individual realizations but rather uses the mean data vector (with the covariance appropriately rescaled by the number of realizations), the above calculation shows that the expectation of $\sigma_\text{red} \sim \sqrt{2/(N_T N_L)}$ is correctly recovered both when using the sample covariance and when using a noiseless covariance. 

It is easy to support these results through numerical experiment; we can use e.g. $N_T = 1$ for simplicity (the scaling with $N_T$ is trivial), $N_r = 20, N_L = 5$ and generate many realizations of $\hat{\bar{\chi}}^2$ either using the diagonal sample covariance or the true theoretical covariance. The resulting distribution of $\hat{\bar{\chi}}^2$ from $100,000$ draws was then determined. The expected variance from Eq. \eqref{eq:varred} matched the observed sample variance from the $100,000$ draws to about one percent. After applying these corrections, we find better agreement between posteriors and chi-squared values derived from the theoretical, noiseless Gaussian covariance and the diagonal sample variance, thus validating our reasoning\footnote{Since we also set cross-covariances between bispectra containing B-modes and bispectra not containing B-modes to zero in the diagonal sample covariance, there is some ambiguity as to what $N_L$ should be in those cases (strictly speaking, an overall rescaling is no longer possible). However, we ignore this difference as its impact on our overall conclusions is small.}.
\section{C: Line-of-sight dependence of Shape Field}\label{sec:los}
In \cite{bakx_eft} it was shown in their Appendix B that shape fields estimated from simulations using the conventional definitions 
\begin{equation}
    \gamma_1^\text{sims} = \frac{I_{11}-I_{22}}{I_{11}+I_{22}}, \quad \gamma_2^\text{sims} = \frac{2I_{12}}{I_{11}+I_{22}} 
\end{equation}
suffer from spurious additional line of sight dependence beyond what is expected based on the flat-sky projection formalism introduced in Section \ref{sec:theory}, which assumes a normalization by a rotationally invariant quantity such as $\text{Tr}(I_{ij})$, i.e.
\begin{equation}
    \gamma_1^\text{theory} = \frac{I_{11}-I_{22}}{I_{11}+I_{22}+I_{33}}, \quad \gamma_2^\text{theory} = \frac{2I_{12}}{I_{11}+I_{22}+I_{33}}. 
\end{equation}
Although \cite{chen_lensing} mentions that this effect appears to be subdominant for practical applications, we reconsider it here in case of the bispectrum. It is most transparent to do this for the case of $B_{\delta \delta E}$ in the collinear limit. For collinear triangles, the anisotropic bispectrum in the theory model depends only on a single unit vector. By imposing statistical isotropy, the only possible angular dependence is thus 
\begin{equation}
    B^\text{ssg}_{ij}(\bb{k}_1,\bb{k}_2,\bb{k}_3) = B^\text{ssg}_{ij}(k_1\hat{\bb{k}},k_2\hat{\bb{k}},k_3\hat{\bb{k}}) = \mathcal{D}_{ij}(\hat{\bb{k}})B_0(k_1,k_2,k_3)
\end{equation}
and as in the case of the power spectrum $P_{\delta E}$, there is only one independent multipole. We stress that this assertion is independent of any dynamics. For the collinear configuration, there is no B-mode, as expected from symmetries. We thus obtain the relations\footnote{Note that this remains valid even if we integrate over the radial bins explicitly, and also regardless of whether the angle $\xi$ is well-defined (which it is not when the wave vectors are exactly collinear).} 
\begin{equation}\label{eq:rat1}
    \bigg[\frac{B_{\delta\delta E}^{20}(k_1,k_2,k_3)}{B_{\delta\delta E}^{00}(k_1,k_2,k_3)}\bigg]^\text{theory} = \frac{5\int_{-1}^1\md \mu \, \frac{1}{2}(3\mu^2-1)(1-\mu^2) \bb{M}_{ij}^E(\hat{\bb{k}})\mathcal{D}_{ij}(\hat{\bb{k}})}{\int_{-1}^1\md \mu\, (1-\mu^2) \bb{M}_{ij}^E(\hat{\bb{k}})\mathcal{D}_{ij}(\hat{\bb{k}})} = \frac{5\int_{-1}^1\md \mu \, \frac{1}{2}(3\mu^2-1)(1-\mu^2)^2}{\int_{-1}^1\md \mu\, (1-\mu^2)^2} = -\frac{10}{7}
\end{equation}
for collinear configurations with $k_1 \approx k_2 + k_3$ and similarly
\begin{equation}\label{eq:rat2}
    \bigg[\frac{B_{\delta\delta E}^{11}(k_1,k_2,k_3)}{B_{\delta\delta E}^{00}(k_1,k_2,k_3)}\bigg]^\text{theory} = \frac{9\int_{-1}^1\md \mu \, \mu(-\mu)(1-\mu^2) \bb{M}_{ij}^E(\hat{\bb{k}})\mathcal{D}_{ij}(\hat{\bb{k}})}{\int_{-1}^1\md \mu\, (1-\mu^2) \bb{M}_{ij}^E(\hat{\bb{k}})\mathcal{D}_{ij}(\hat{\bb{k}})} = \frac{9\int_{-1}^1\md \mu \, \mu(-\mu)(1-\mu^2)^2}{\int_{-1}^1\md \mu\, (1-\mu^2)^2} = -\frac{9}{7}.
\end{equation}
These equalities do not strictly hold for the bispectra derived from the measured shape fields $\gamma_{1,2}^\text{sims}$, since the normalization $I_{11} + I_{22}$ is not rotationally invariant.
\begin{figure}
    \centering
    \includegraphics[width=0.5\linewidth]{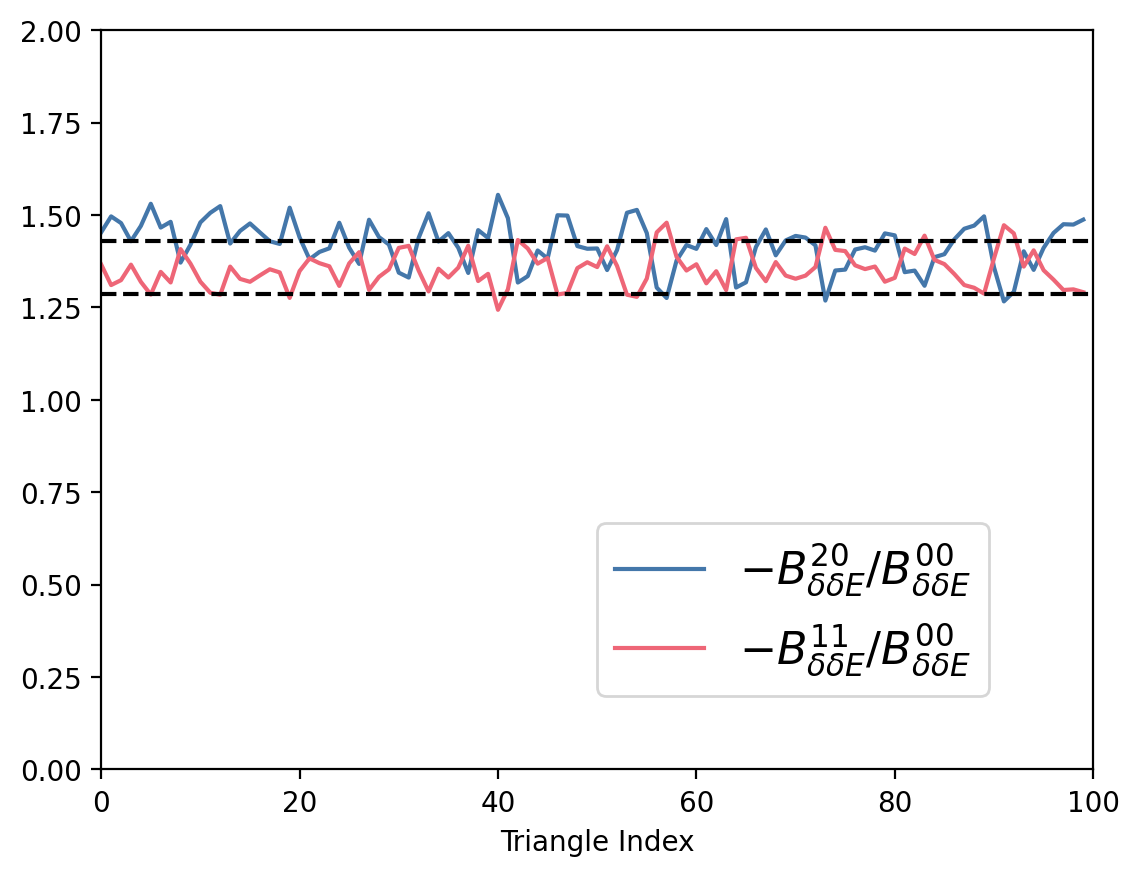}
    \caption{Ratios from Eqs. \eqref{eq:rat1} and \eqref{eq:rat2} as a function of triangle index, where we select all open triangle bins with $k_1\geq k_2\geq k_3$ and shortest side longer than $k_{\text{min}} = 0.06$. The long-dashed lines indicated the theoretically expected values of $9/7$ and $10/7$ from Eqs. \eqref{eq:rat1} and \eqref{eq:rat2}.}
    \label{fig:ratios}
\end{figure}

Figure \ref{fig:ratios} shows the ratio of measurements of the $(\ell_1,\ell_2) = (0,0)$ and $(\ell_1,\ell_2) = (2,0)$ moments of the collinear bispectrum. We also consider the ratio of the lowest order multipole to the other quadrupole, that is $(\ell_1,\ell_2) = (1,1)$. For these wavenumber configurations, we see that the ratios are roughly scale-independent and in accordance with their theoretically predicted values (given the discreteness effects we already encountered in the main text, we do expect to see some deviations at the level shown in the figure). We thus conclude that this discrepancy is not a source of systematics for our purposes.  

\bibliographystyle{JHEP}
\bibliography{mybibliography}
\end{document}